\documentclass[aps,prstab,reprint,groupedaddress,showpacs,floatfix]{revtex4-1}

\usepackage{amsmath}
\usepackage{bm}
\usepackage[caption=false,lofdepth,lotdepth]{subfig}
\usepackage{graphicx}
\usepackage{dcolumn}

\begin{document}

\title{Laserwire at the Accelerator Test Facility 2 with Sub-Micrometre Resolution}

\author{L.~J.~Nevay}
\email{laurie.nevay@rhul.ac.uk}
\author{S.~T.~Boogert}
\author{P.~Karataev}
\author{K.~Kruchinin}
\affiliation{John Adams Institute at Royal Holloway, University of London, Egham, TW20 0EX, United Kingdom}

\author{L.~Corner}
\author{D.~F.~Howell}
\author{R.~Walczak}
\affiliation{John Adams Institute at University of Oxford, Denys Wilkinson Building, Oxford OX1 3RH, United Kingdom}

\author{A.~Aryshev}
\author{J.~Urakawa}
\author{N.~Terunuma}
\affiliation{KEK, 1-1 Oho, Tsukuba, Ibaraki 305-0801, Japan}

\date{\today}

\begin{abstract}
A laserwire transverse electron beam size measurement system has been developed and operated 
at the Accelerator Test Facility 2 (ATF2) at KEK. Special electron beam optics were developed 
to create an approximately 1\,$\times$\,100\,$\mu$m (vertical\,$\times$\,horizontal) 
electron beam at the laserwire location, 
which was profiled using 150\,mJ, 71\,ps laser pulses with a wavelength of 532\,nm. 
The precise characterisation of the laser propagation allows the non-Gaussian laserwire 
scan profiles caused by the laser divergence to be deconvolved. 
A minimum vertical electron beam size of 
1.07~$\pm$~0.06~(\textit{stat.})~$\pm$~0.05~(\textit{sys.})\,$\mu$m 
was measured. A vertically focussing quadrupole just before the laserwire was 
varied whilst making 
laserwire measurements and the projected vertical emittance was measured to be
\mbox{82.56~$\pm$~3.04\,pm\,rad}.
\end{abstract}


\maketitle


\section{Introduction \label{sec:introduction}}

For future linear electron-positron colliders such as the Compact Linear Collider 
(CLIC)~\cite{Clic2012} and the International Linear Collider (ILC)~\cite{Ilc2007}, 
measurement of the particle beam emittance is essential to achieve and maintain 
the required nanometre-level final focus beam sizes if their target luminosity is 
to be reached. The emittance is typically measured by measuring the transverse profile of 
the beam at several points in the lattice with a different betatron phase advance, or 
by having a single beam size measurement location and changing the strength of an upstream 
quadrupole. 
Two commonly used methods at electron accelerators to measure the transverse beam profile are 
optical transition 
radiation (OTR) screens that image the beam directly~\cite{Ross2002} 
and wire-scanners~\cite{Hayano2000}. However, both of these devices are destructive and 
themselves suffer damage from 
high charge density beams and therefore are not suitable to make measurements at 
the full beam energy or bunch charge as would be required for continuous monitoring 
and tuning of 
the accelerator optics in future linear colliders~\cite{Tenenbaum1999}. 

A laserwire is a beam profile monitor based on Compton scattering of laser photons
from the electrons or positrons in the particle beam~\cite{Agapov2007}.
With high energy particle beams, 
the scattered photons have a high energy and travel nearly parallel to the particle beam
and can be detected after a bend in the beamline, that deflects the charged particle beam. 
As the laser focus is scanned transversely across 
the particle beam, the rate of Compton-scattered photons is modulated yielding a laserwire
scan. With knowledge of the laser size at its focus, the laserwire scan can be deconvolved to 
give the electron beam profile. As the Compton cross-section is very small, a high power
pulsed laser source must be used and only a small fraction of the bunch particles
are scattered.

\begin{figure*}
\normalsize
\includegraphics[width=1.0\textwidth]{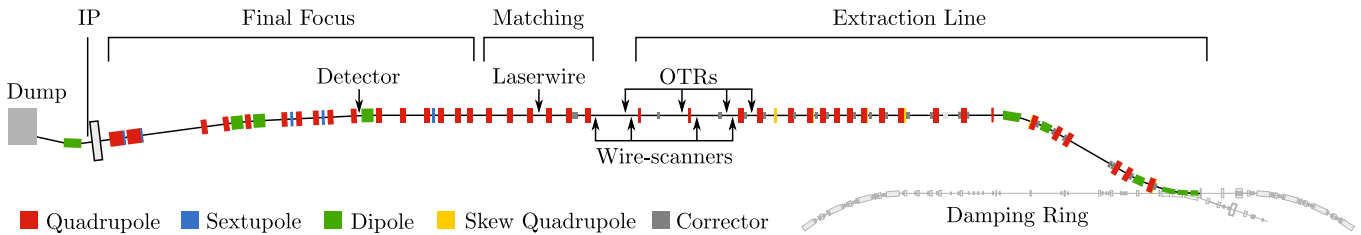}
\caption{\label{fig:extline} Schematic of the ATF2 extraction line showing the location of 
the laserwire system as well as the detector located immediately behind the first dipole 
magnet after the laserwire interaction point. The wire-scanners and optical transition 
radiation (OTR) monitors are also shown at the end of the extraction line section. All 
quadrupoles and sextupoles in the matching and final focus sections are on individual 3-axis
mover systems.}
\hrulefill
\end{figure*}

A number of laserwire beam profile monitors have been demonstrated such as those at 
Stanford Linear 
Collider (SLC)~\cite{Ross2003,Alley1996} and at PETRAII in DESY~\cite{Aumeyr2010}. 
In the case of the laserwire at SLC, an ultraviolet wavelength laser was used to achieve
micrometre size laserwire scans. A focussing geometry consisting of both transmissive 
and reflective optics was chosen to compensate for spherical aberrations, allowing 
the micrometre size focussed laser spot sizes to be
achieved. However, the reflective geometry prevents measurement of the focussed laser 
spot directly and therefore calibration as measuring the focussed spot will 
intercept the incoming laser beam. It is also not possible to measure the 
laser pulse energy directly and strictly limits the scanning range.
Alternatively, transmissive focussing optics allow direct access to the laser focus
for measurement and therefore laserwire calibration. It also affords a greater 
scanning range and allows the laser pulse energy to be measured after the interaction
point for normalisation purposes. As the minimum focussed spot size of the 
laser beam (and therefore the resolution 
of the laserwire) is limited by the wavelength of light, a wavelength less than
1\,$\mu$m must be used to measure a 1\,$\mu$m electron beam. Wavelengths below 300~nm 
limit the choice of optical materials due
to absorption and also necessitate a higher power laser system for the same delivered peak 
laser power. A good compromise is the use of a visible wavelength laser with 
transmissive optics~\cite{Boogert2010}. 

The ATF is a prototype damping ring~\cite{Honda2004} with an extraction and dump 
line where our first laserwire system was installed~\cite{Boogert2010}, which measured
a minimum electron beam size of 
\mbox{4.8~$\pm$~0.3\,$\mu$m}. However it was 
observed that Rayleigh range effects were present and the model used to describes this 
requires knowledge of the horizontal electron beam size. 
Without this knowledge, the model left
some ambiguity about the size of the electron beam. The extraction line was significantly
upgraded to create a prototype final focus system for future linear colliders,
called ATF2~\cite{Report2006,White2014}. This paper presents results from 
the upgraded ATF laserwire system that aims to achieve micrometre sized 
transverse profiles using 
a visible wavelength laser system. Precise laser characterisation and horizontal laserwire
scans allow the more detailed model to be used and accurately measure both the horizontal and
veritcal electron beam sizes.

\section{Setup \label{sec:setup}}

\begin{figure}
\includegraphics[width=8.6cm]{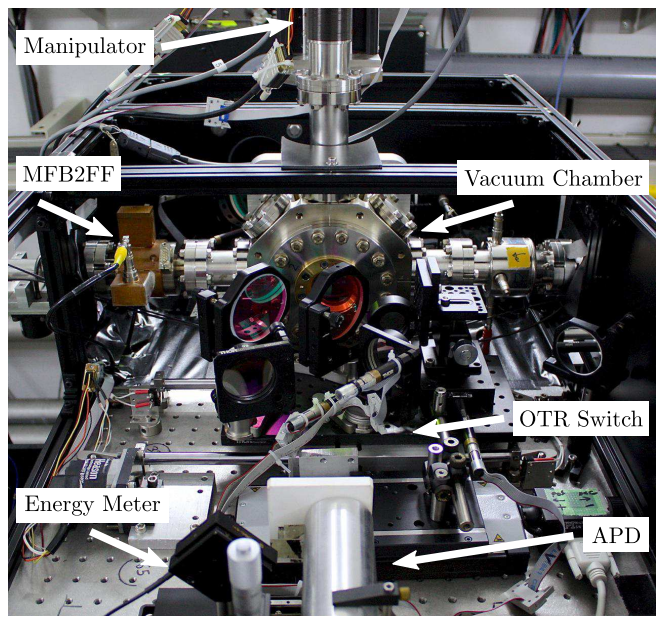}
\caption{\label{fig:lwsetup} Photograph of the laserwire installation in the ATF2 beam line. 
The electron beam travels from right to left and the laser beam enters behind the vacuum 
chamber and exits towards the reader. The manipulator for the optical transition
radiation (OTR) and alignment screen 
can be seen on top of the vacuum chamber. The avalanche photodiode (APD) used for timing 
and the laser pulse energy meter can be seen in the foreground. The high resolution 
cavity BPM MFB2FF is also shown attached to the laserwire vacuum chamber. The small optical
breadboard (OTR Switch) allows switching between the high power laser path for laserwire 
and the low intensity OTR path.}
\end{figure}

To demonstrate the desired micrometre size profiles, the laserwire system was positioned
in the matching section of the ATF2 extraction line at a virtual image point of the final
focus in the vertical as shown in Figure~\ref{fig:extline}. A photograph of the 
experimental setup in the ATF2 beam line is shown in Figure~\ref{fig:lwsetup}. 
An overview of the various subsystems and their upgrades from the laserwire installation
at the ATF are described in the following sections. 


\subsection{ATF2 Electron Beam Optics}

\begin{table}
\caption{\label{tab:atfparams} ATF2 Parameters}
\begin{tabular}{l c c l}
\hline \hline
Parameter             & Symbol             & Value               & Units \\ \hline
Beam Energy           & E                  & 1.30                & GeV \\
Horizontal emittance  & $\gamma\epsilon_{x}$  & 4\,$\times$\,$10^{-6}$ & m\,rad \\
Vertical emittance    & $\gamma\epsilon_{y}$  & 4\,$\times$\,$10^{-8}$ & m\,rad \\
Bunch repetition rate & $f_{\rm{bunch}}$     & 3.12                & Hz   \\ 
Bunch length          & $\sigma_{ez}$       & $\sim$\,30           & ps   \\ 
Electrons per bunch   & $N_{e}$             & 0.5\,-\,10\,$\times$\,10$^{9}$ & e$^{-}$\\
Fractional momentum spread & $\Delta p/p$   & 0.001             & \\ \hline\hline
\end{tabular}
\end{table}

\begin{figure}
\includegraphics[width=8.6cm]{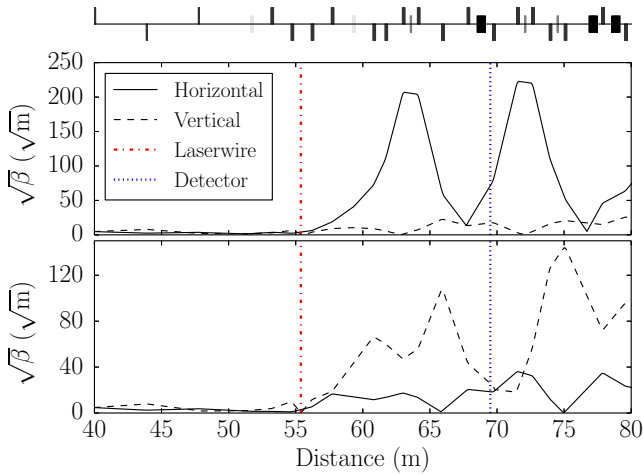}
\caption{\label{fig:optics}Electron beam amplitude functions for the end of the extraction line, 
matching section and beginning of the final focus section. These are 
shown for normal ATF2 operation (\textit{top}) and for laserwire operation 
(\textit{bottom}). The laserwire and laserwire detector locations are shown 
by (\textit{red}) dot-dashed and (\textit{blue}) dotted vertical lines
respectively.}
\end{figure}

\begin{figure}
\includegraphics[width=8.6cm]{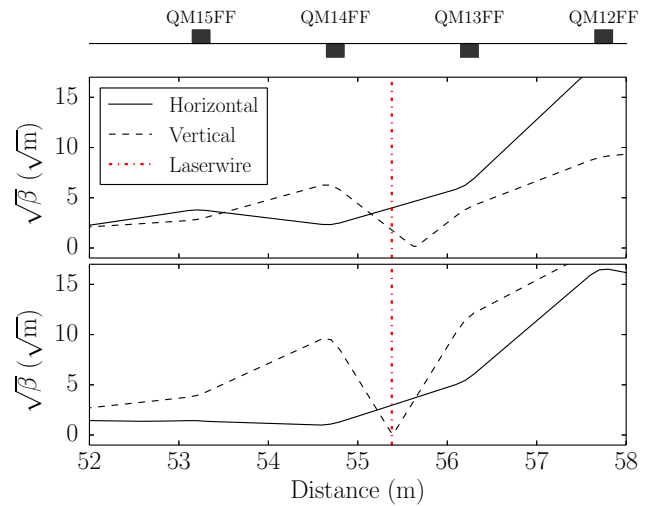}
\caption{\label{fig:opticszoom}Electron beam amplitude functions about the 
laserwire interaction point
for normal ATF2 operation (\textit{top}) where the vertical waist is located at the MFB2FF
cavity BPM, and for laserwire operation (\textit{bottom}), where the waist is moved to
the laserwire location.}
\end{figure}

A summary of the electron beam parameters is given in Table~\ref{tab:atfparams}. During
laserwire operations, a specially developed set of electron beam optics was used to 
minimise the vertical beam size at the laserwire interaction point (LWIP). 
In Figure~\ref{fig:optics}, both sets of electron beam optics are shown for the extraction
line, matching section and beginning of the final focus. With the 
normal ATF2 settings, a vertical waist exists $\sim$\,20\,cm downstream from the LWIP 
at the location of the MFB2FF cavity BPM (CBPM). The laserwire electron beam optics were
designed to move the waist to the LWIP and reduce the vertical electron beam 
size further. Additionally, with the normal ATF2 optics, the horizontal amplitude 
function $\beta_{x}$ expands in the final focus section to $\beta_{x}~>~40$\,km as
can be seen in the top part of Figure~\ref{fig:optics}, 
which intentionally collimates the electron beam using the beam pipe. This however, 
produces a large background for the laserwire detector, and so an effort was made to 
significantly reduce $\beta_{x}$ in the final focus section.
As expected, reducing $\beta_{x}$ increases $\beta_{y}$, 
but this does not generate comparable background levels as 
the vertical emittance is much lower than the horizontal emittance. 
Figure~\ref{fig:opticszoom} shows $\beta_{x}$ and $\beta_{y}$ around the laserwire location, 
where the shifted vertical waist is clear. 

Using the laserwire electron beam optics, \mbox{$\beta_{x}^{*}~=~8.822$\,m} and 
\mbox{$\beta_{y}^{*}~=~15.625$\,mm} at the LWIP ($*$ denotes the value of functions at the LWIP), 
which combined with the nominal emittance values gives 
a predicted electron beam size
of \mbox{0.495\,$\mu$m\,$\times$\,117\,$\mu$m} (vertical\,$\times$\,horizontal).
Apart from the amplitude functions $\beta_{x,y}$, the electron beam size $\sigma_{e}$ also
depends on the dispersion at the LWIP as described by

\begin{equation}
\label{eq:ebeamsize}
\sigma_{e} = \sqrt{ \epsilon\,\beta + D^{2}\left(\frac{\Delta p}{p}\right)^{2}}
\end{equation}

\noindent where $\epsilon$ is the geometric emittance, $\beta$ the beta amplitude function, 
$D$ the 
dispersion, and $\Delta p/{p}$ is the fractional momentum spread of the 
beam. Although the horizontal and vertical dispersion at the LWIP are nominally zero, 
there will be a finite amount due to residual dispersion from beam misalignment in the 
extraction line quadrupoles or residual 
$x$-$y$ coupling. Dispersion and coupling are measured and corrected
using the Flight Simulator software~\cite{White2008}, which measures the electron beam
trajectory using the high resolution CBPM system~\cite{Kim2012} as a function 
of electron beam energy. The energy is modulated by adjusting the damping ring rf frequency, and
the calculated coupling and dispersion corrections applied using four upstream skew 
quadrupoles in combination.

The CBPM system provides high resolution position measurement at 45 locations
through the extraction line, matching section and final focus section of the ATF2. The
majority of the CBPMs are mounted to the pole faces of the quadrupoles in the matching 
and final focus sections, with the remainder at other points in the extraction line.
There are CBPMs in the quadrupoles before and after the LWIP, however the CBPM afterwards
is on the far side of the quadrupole, and so the trajectory cannot be treated as ballistic
between the two.  A high resolution CBPM, MFB2FF, is attached to the 
laserwire vacuum chamber and moves with it during laserwire scans.  MFB2FF has a 
typical resolution of 70\,nm at the bunch charge used during laserwire operations
over a limited range of \textless~100\,nm~\cite{Kim2012}. The scanning range of the 
laserwire exceeds this range and the mechanical offset and tilt of MFB2FF in 
relation to the laserwire vacuum chamber introduced $x$-$y$ coupling and 
degrades the resolution. Therefore, the electron beam position from MFB2FF
was not suitable for spatial jitter subtraction during laserwire operation. 

Although the CBPMs near the LWIP could be used for spatial jitter subtraction 
at the LWIP, those around the laserwire are configured for a large dynamic 
range at the expense of resolution and have a typical resolution of 200\,$\mu$m.  
Given the approximately 1\,$\mu$m vertical electron beam sizes with the laserwire 
electron beam optics, spatial jitter subtraction was not possible. 
As the laserwire averages over multiple pulses, the approximate
electron beam jitter of $\sim$\,0.2\,$\sigma$ at the ATF2 will contribute a systematic 
increase of approximately 1.9\,\% to the measured electron beam size.

\subsection{Laserwire Interaction}

\begin{figure}
\includegraphics[width=6.88cm]{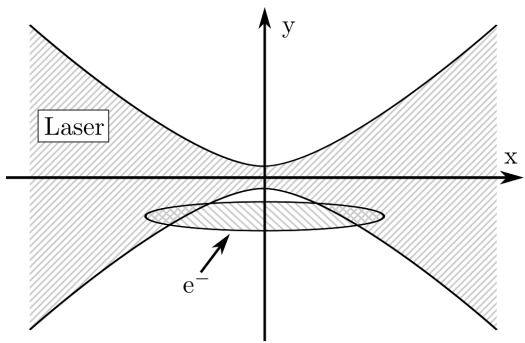}
\caption{\label{fig:aspectratio}Schematic of the laser focus showing its interaction with 
the high aspect ratio electron beam even when the laser focus is displaced from the 
electron beam. The vertical axis is expanded in scale compared to the horizontal to make
the overlap clearer.}
\end{figure}

A laserwire scan consists of the measurement of the Compton-scattered photon
rate for different laser focus positions and is therefore the convolution 
of the transverse density functions of the laser beam and the electron beam
in the axis of the scan.  
If both the electron and laser beams have Gaussian density functions, the 
convolution also has a Gaussian form.  However, unlike the wire in a 
traditional wire-scanner, the laser beam size is not constant and expands on either 
side of the focus. The length scale of this is described by the Rayleigh range of
the laser beam, which is the distance from the focus until the beam expands to 
twice its area. In the case where the electron beam 
size along the laser propagation axis is much smaller than the Rayleigh range, the
laser width can safely be assumed to be constant across the electron beam width and
the laserwire scan can be easily deconvolved independently of the horizontal size 
of the electron beam. However, in 
the case where the electron beam size is comparable to or greater than the Rayleigh 
range, the divergent laser beam interacts with the electron beam 
even when the laser focus is significantly displaced from the centre of the 
electron beam as shown schematically in Figure~\ref{fig:aspectratio}. To 
deconvolve the laserwire scan and measure the electron beam size, the form of
the scan must be derived.

\begin{widetext}
\begin{equation}
\label{eq:lwoverlapintegral}
N_{c}(\Delta_{x},\Delta_{y}) = 
\frac{P_{l}\, N_{e} \, \lambda \,\sigma_{c}}{hc^{2}} \frac{1}{2\pi\sigma_{ex}}
\int_{-\infty}^{+\infty} \frac{1}{\sqrt{\sigma_{ey}^{2} + \sigma_{l}(x-\Delta_{x})^{2}}}\, 
\mathrm{exp} \left\{ - \frac{x^{2}}{2\sigma_{ex}^{2}} - 
\frac{\Delta_{y}^{2}}{2[\sigma_{ey}^{2} + \sigma_{l}(x-\Delta_{x})^{2}]} \right\}
\,\mathrm{d}x
\end{equation}
\end{widetext}

As the Compton cross-section is constant for a given laser wavelength, electron beam
and collision geometry, the rate of Compton-scattered photons is determined by 
calculating the luminosity of the laser-electron
beam collision for different offsets of the laser beam from the electron beam. The 
luminosity is calculated using the overlap integral between the 
four dimensional density functions of the laser and electron beams.
It has been shown in~\cite{Agapov2007} that by assuming Gaussian density distributions 
for the laser and electron beams and in the case where the laser pulse
length is much longer than the electron bunch length, that the laser distribution
can be treated as a static target and integration over three of the dimensions can be 
solved analytically yeilding Equation~\ref{eq:lwoverlapintegral}. Here, 
$P_{l}$, $N_{e}$, $\lambda$ and $\sigma_{c}$ are the peak laser 
power, number of electrons per bunch, the wavelength of the laser and the Compton 
cross-section respectively. This equation describes the number of Compton-scattered 
photons $N_{c}$ as function of $\Delta_{x}$ and $\Delta_{y}$; the horizontal and 
vertical displacement of the laser focus from the centre of the electron beam.
The vertical and horizontal sizes of the electron beam are $\sigma_{ey}$ and $\sigma_{ex}$ 
and the vertical laser beam size is 
\mbox{$\sigma_{l}(x-\Delta_{x})$}. The width of a laser beam is conventionally defined by $w$, 
however, we use $\sigma$ (where \mbox{$w~=~2\sigma$}) for compatability with electron beam sizes. 
The propagation of a focussed multimode laser beam
is given by 

\begin{equation}
\label{eq:lsigma}
\sigma_{l}(x-\Delta_{x}) = 
\sigma_{lo}\,\sqrt{1 + \left[ \frac{(x-x_{\sigma lo}-\Delta_{x})} {x_{R}} \right]^{2}}
\end{equation}

\noindent where $\sigma_{lo}$ is the minimum size of the laser at its focus and 
$x_{\sigma lo}$ is the location of the focus. The Rayleigh range $x_{R}$ is 
given by

\begin{equation}
\label{eq:rayleighrange}
x_{R} = \frac{\pi\,(2\,\sigma_{lo})^{2}}{M^{2} \lambda}
\end{equation}

\noindent where $M^{2}$ is a linear scaling parameter describing
the spatial quality of the laser beam with respect to one with a pefectly Gaussian transverse
intensity profile and can be measured~\cite{Johnston1998}. In the case 
where \mbox{$\sigma_{ex}~\ll~x_{R}$}, 
and the laser focus is aligned to the electron beam centre ($\Delta_{x}~=~0$), 
Equation~\ref{eq:lwoverlapintegral} simplifies to

\begin{eqnarray}
N_{c}(\Delta_{y}) = 
\frac{P_{l}\,N_{e}\,\lambda\,\sigma_{c}}{hc^{2}}
\frac{1}{(2\pi)^{(3/2)}\sqrt{\sigma_{ey}^{2} + \sigma_{lo}^{2}}}\,\times \qquad\quad\nonumber
\\
\mathrm{exp}\left[-\frac{\Delta_{y}^{2}}{2\,(\sigma_{ey}^{2} + \sigma_{lo}^{2})}\right]
\end{eqnarray}

\noindent which has the form of a Gaussian. In this case, the laserwire vertical scan is 
independent of the horizontal beam size, and with knowledge of $\sigma_{lo}$, the 
$\sigma$ of the Gaussian laserwire scan can be analytically deconvolved to give 
the size of the electron beam, $\sigma_{ey}$.

However, in the case where $\sigma_{ex}$ is greater than or comparable to $x_{R}$, 
Equation~\ref{eq:lwoverlapintegral} must be used with the measured laser propagation
and horizontal electron beam size $\sigma_{ex}$. This 
presents a significant limit on the use of a laserwire as a beam diagnostic and 
especially so at the ATF2 where the horizontal electron beam is expected to be 
$\sim$\,100\,$\mu$m, which will be comparable or greater than the Rayleigh range of the laser.
The natural divergence of the laser beam cannot be avoided and is
dictated by the wavelength used. The laserwire was operated with the assumption 
that horizontal measurements would be concurrently available from the OTR monitor 
installed at the LWIP~\cite{Karataev2011}. 

The Compton-scattered photons from the laserwire have a broad spectrum with a maximum 
energy of

\begin{equation}
h\,\nu_{\rm{max}} = E \left(\frac{2 \xi}{1 + 2 \xi}\right)
\label{eq:comptonmaxenergy}
\end{equation}

\noindent where $E$ is the energy of the electron beam and $\xi$ the normalised photon 
energy in the electron rest frame \mbox{($\xi~=~\gamma\,h\,\nu\, / m_{e} c^{2}$)}. 
In the case 
of the ATF2, with a laser wavelength of 532\,nm, the maximum Compton-scattered photon 
energy is 29.4\,MeV. The Compton cross-section is also dependent on the energy 
of the electron beam and the laser wavelength and in this case, the total cross-section 
averaged over all scattered energies is 
6.5\,$\times$\,$10^{-24}$\,m$^{-2}$.

\subsection{Laser System and Optical Transport}

The laser system consists of a Q-switched neodymium-doped yttrium aluminium garnet (Nd:YAG) 
amplifier seeded by a 357\,MHz mode-locked 
oscillator. The laser oscillator is stabilised and locked to an external signal generator at 
approximately 357\,MHz by means of a piezo actuator on one of the oscillator mirrors, with a
typical temporal jitter of \textless~2\,ps. The signal generator is manually adjusted to match 
the ATF2 master oscillator frequency at the start of each operation period, as 
the ATF2 frequency is varied slightly according to the annual 
expansion cycle of the damping ring, and so the laserwire system must match it. A 10\,MHz
reference signal relayed between the ATF2 master oscillator and the laserwire signal
generator ensures a stable phase relationship between the laser pulses and the electron
bunches. Two electro-optic 
modulators are used to isolate a single laser pulse from the oscillator, which is 
then amplified first in a regenerative amplifier followed by passage through a spatial 
filter and two single-pass linear amplifiers. The 1064\,nm wavelength light is then 
frequency doubled in a beta-Barium Borate ($\beta$-BaB$_{2}$O$_{4}$, BBO) 
crystal providing $\sim$\,150\,mJ pulses with a wavelength of 532\,nm and a duration of 
$\sigma_{\tau}~=~70.8~\pm~0.6$\,ps at 3.12\,Hz, the repetition rate of the ATF2.

The necessary trigger signals for the laser system are 
derived from the ATF2 damping ring extraction kicker thyratron charge and fire signals, which 
happen 1\,ms before the extraction of the beam from the damping ring and 
the extraction itself respectively. 
The trigger for the regenerative amplifier and linear amplifiers are 
independently controllable providing a large range of output pulse energy levels
of the laser system. The locally generated 357\,MHz signal is first passed through a 
voltage controlled phase shifter allowing all of the laser trigger signals, that are 
created multiple digital counter and delay generators, to be adjusted with respect 
to the electron beam arrival time whilst still maintaining their respective phase 
and timing relationships. 

The laser system is mounted on an optical table in a temperature controlled lab on 
top of the accelerator concrete shielding blocks. A 10\,cm diameter hole in both the 
table and the shielding blocks allows the laser beam to be transported in free space 
using mirrors into the accelerator environment. An automated mirror insert in the lab 
allows the laser beam path to be switched to a laser diagnostic
line that consists of a series of relay mirrors providing exactly the same optical path 
length as to the laserwire lens beside the LWIP ($\sim$\,8.4\,m). A further mirror insert 
allows a 25\,mW continuous-wave laser to be used for alignment purposes. 

\subsection{Interaction Point}

\begin{figure}
\includegraphics[width=6.9cm]{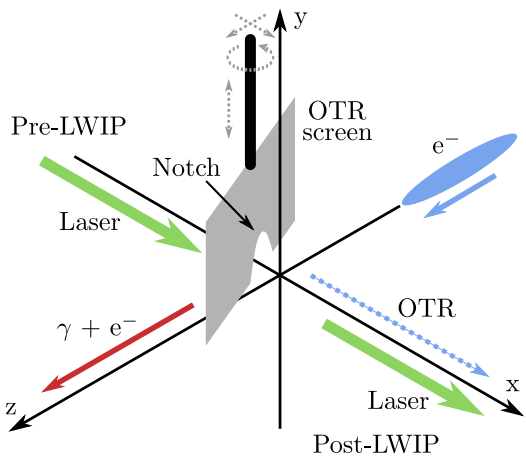}
\caption{\label{fig:lwaxes}Schematic of the beam geometry at the laserwire 
interaction point, including the OTR screen at 45$^{\circ}$ to the electron beam 
direction, incoming electron bunch, outgoing, OTR path, laser beam path, 
and Compton-scattered photons ($\gamma$).}
\end{figure}

After the laser is transported to the laserwire interaction point, it is directed into 
a custom-made vacuum chamber with high damage threshold vacuum windows on either side to
allow the laser beam to enter and exit. The vacuum chamber can be moved $\pm$~3\,mm in 
both the horizontal $x$ and vertical $y$ axes. 
A kinematic lens mount attached to the vacuum chamber is used to mount the laserwire lens, 
which allows precise control of the lens separation from the vacuum window as well as its angle. 
This is imperative as the vacuum window is an integral part of the lens optical design.
The laserwire lens ($f~=~56.6$\,mm) consists of two radiation-hard fused silica elements that are 
designed to correct geometric aberrations. The high radiation environment of the accelerator
permits only fused silica to be used and therefore without different lens materials, chromatic
aberrations cannot be corrected. The Nd:YAG laser source provides
narrow bandwidth laser pulses that are easily accomodated by the 2\,nm acceptance bandwidth of
the lens negating any chromatic aberrations. By moving the vacuum chamber, the attached lens
and therefore the laser focus also move. Optical position encoders provide 50\,nm accuracy on 
the chamber position measurement. The coordinate axes of the interaction 
point are shown in Figure~\ref{fig:lwaxes}. 

A screen for both OTR and alignment is mounted on a vacuum manipulator arm that enters the 
vacuum chamber through the top access port. Manual micrometers allow the manipulator arm and 
therefore the screen to be moved in the $x$ and $z$ axes, while motorised actuators 
control the angle of the screen $\theta_{\rm{OTR}}$ and its vertical position 
in the $y$ axis. 

\begin{figure}
\includegraphics[width=8.43cm]{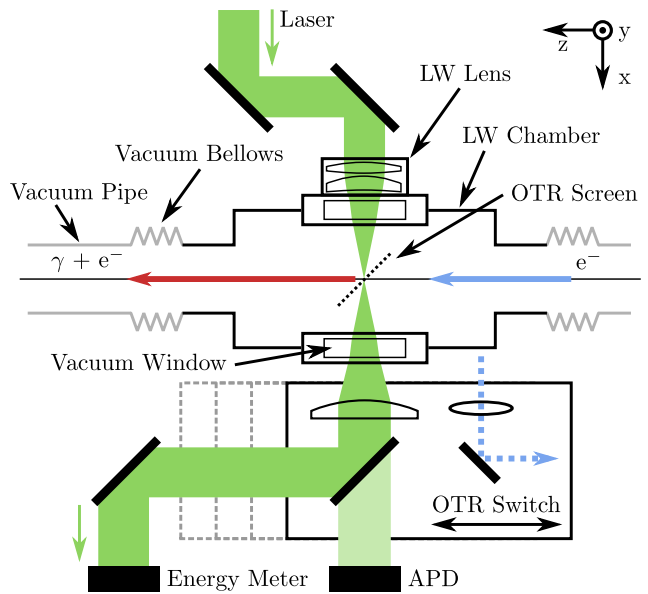}
\caption{\label{fig:lwipplan}Schematic of the laserwire (LW) interaction point in plan view 
showing the lens and vacuum windows attached to the vacuum chamber, the laser beam 
path (\textit{green}) and post-LWIP optical switch for the optical transition radiation (OTR). 
The laser beam enters at the top of the diagram and is absorbed in the energy meter. The
avalanche photodiode (APD) is used for timing purposes.}
\end{figure}

After the interaction point (post-LWIP), the laser beam exits the vacuum chamber through the 
vacuum window and is directed by two mirrors onto a laser energy meter. A plano-convex 
lens is used to bring the laser beam inside the active area 
of the energy meter. The post-LWIP optics are required to deal with the safe disposal 
of gigawatt peak power laser pulses, but also to image OTR, which is
$\sim$\,10$^{10}$ lower in intensity. To 
accomodate this, two separate switchable optical paths are used. 
Mirrors for each optical path are fixed on to a small optical breadboard that is 
mounted on top of a translation stage. Figure~\ref{fig:lwipplan} shows the layout
schematically.

An avalanche photodiode (APD) is used to simultaneously detect the laser light when 
strongly attenuated and a combination of OTR, optical diffraction radiation (ODR) and 
reflected synchrotron 
radiation (SR)~\cite{Karataev2011} from the electron beam, allowing synchronisation of both. 
The first post-LWIP high-reflectivity dielectic-coated mirror is used to attenuate the 
laser pulses without affecting the broadband OTR.

\subsection{Detector}

The laserwire detector is placed after the BH5X dipole magnet in the ATF2 lattice, 
which is the first bend after the LWIP and constitutes a bend 
of 2.927$^{\circ}$. The box-shaped vacuum pipe in the dipole has an 
aluminium window 
26\,mm in diameter and 200\,$\mu$m in thickness at the end that allows the Compton-scattered 
photons from the laserwire to be detected. 

The detector consists of a \mbox{4\,$\times$\,4\,$\times$~0.6\,cm$^{3}$} ($x$\,$\times$\,$y$\,$\times$\,$z$) lead 
sheet that acts as a converter of photons to electron-positron pairs, followed by a 
\mbox{4\,$\times$\,4\,$\times$\,5\,cm$^{3}$} block of SP15 Aerogel. The Aerogel acts as a 
Cherenkov radiator for the electron-positron pairs and the Cherenkov light is guided 
in a light tight pipe, internally coated with aluminiumised Mylar, to a shielded 
photo-multiplier tube (PMT) out of the accelerator plane. The detector linearity was 
verified in~\cite{Boogert2010}. Synchrotron radiation background was expected to be 
negligible as the synchrotron photon energy at the peak of its spectrum is $\sim$\,0.3\,keV, which 
is insufficient to generate electron-positron pairs in the lead converter plate.

\subsection{Data Acquisition System \label{ssec:daq}}

The data acquisition system is based around Experimental Physics and Industrial Control 
System (EPICS) database software~\cite{Epics2014}. This provides an easily extendable common 
interface level for all devices that are part of the experimental system as well as a 
graphical user interface using the Extensible Display Manager (EDM) and Python software 
for control, data storage and data analysis. Individual devices are controlled 
through LabView or C software directly, 
which monitor command variables in the EPICS database and publish data and 
measurements to other variables. A suite of Python programs provides high-level control 
of the laser system and laserwire experiment. Data is recorded from all devices 
each machine cycle at 3.12\,Hz as well as data from the CBPM system and other 
ATF2 beam instrumentation~\cite{Kim2012,Alabau2012}.

\section{Results \label{sec:results}}

The laserwire was operated in a series of experimental shifts during January and February 
of 2013 and the results from these operation periods are presented in the following sections.
The detector background level, laser propagation and electron beam properties were all 
characterised before performing laserwire scans and are presented separately. The 
laser and electron beam alignment procedure developed is also detailed. The collision 
data and its analysis are then described in subsequent sections.

\subsection{Detector Background}

The detector background level was measured both during and before accelerator operation. 
The ADC value was recorded with nothing attached to the ADC; when connected to the 
unpowered detector; when the detector was 
powered but there was no electron beam; and lastly during accelerator operations with no
laser beam at the LWIP. These measurements represent the ADC pedestal and noise; the 
electrical pick up of the signal cables; the detector dark current; and the detector background
level cumulatively and respectively. These results are summarised in 
Table~\ref{tab:detectorbackgrounds}.

\begin{table}[h]
\caption{\label{tab:detectorbackgrounds}Detector background levels before and during 
accelerator operation with an electron bunch population of 0.20~$\pm$~0.02\,$\times$\,10$^{9}$\,e$^{-}$.}
\newcolumntype{d}[1]{D{.}{.}{#1} }
\begin{tabular}{ l r r d{-1} }
\hline \hline
State                 & \multicolumn{3}{c}{ADC Counts} \\ \hline
ADC only              & \quad 406.7&$\pm$&2.9   \\
Detector connected    & \quad 368.0&$\pm$&48.9  \\
Detector powered      & \quad 366.7&$\pm$&49.0  \\
Accelerator operation & \quad 741.2&$\pm$&122.5 \\ \hline\hline
\end{tabular}
\end{table}

The ADC alone has a pedestal with a low level of noise. When connected to the laserwire
detector, the pedestal is affected and the noise is significantly greater. The operation
of the detector has a negligible increase in the measured noise. The electrical noise
is most likely due to electrical pick-up of the extraction kicker, which uses high
voltage signals and can be readily observed on most electronics in the vicinity. 
This information was used to subtract the pedestal from the detector signal before 
each experimental shift.

\subsection{Laser Characterisation}

To accurately deconvolve the laserwire scans, precise knowledge of the laser beam 
propagation is required. This is accomplished by measuring the $M^{2}$ of the 
laser and the input laser beam profile to the laserwire lens, which can be used in 
combination to calculate the laser propagation at the LWIP. As the laser 
profile is affected by the passage through various components of the laser system, the 
slight change in alignment often necessary to maintain the required laser output power
necessitates repeating the $M^{2}$ measurement. The laser diagnostic line was 
used to measure the $M^{2}$ and input beam profile in the 
lab without requiring access to the accelerator. Various reflective beam splitters
were used in the diagnostic beam line to reduce the intensity to 
within the dynamic range of the laser beam profiler, allowing accurate measurement 
of the $M^{2}$ at the maximum output level of the laser system as is the case 
during laserwire operation.

\begin{figure}
\includegraphics[width=4.3cm]{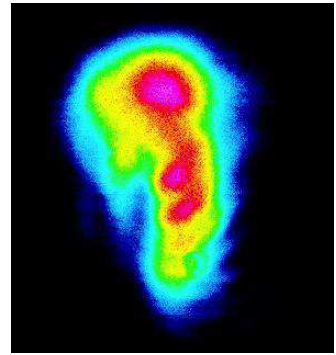}
\caption{\label{fig:lprofile}Input laser profile measured at the end of the
laser diagnostic line with telescope actuator at 12.5\,mm.}
\end{figure}

Previous studies have shown that an input beam size of 
\mbox{$4\sigma$~=~10~-~14\,mm} 
is required on the laserwire lens to produce the smallest possible focussed 
spot size~\cite{Boogert2010}. This size range allows the largest possible input beam
size without incurring either diffraction effects due to the fixed aperture of the lens, or
possible geometric aberrations. The input laser beam profile to the laserwire 
lens was measured at the end of the laser diagnostic line 
in the lab and an example profile is shown in Figure~\ref{fig:lprofile}.

\begin{figure}
\includegraphics[width=8.6cm]{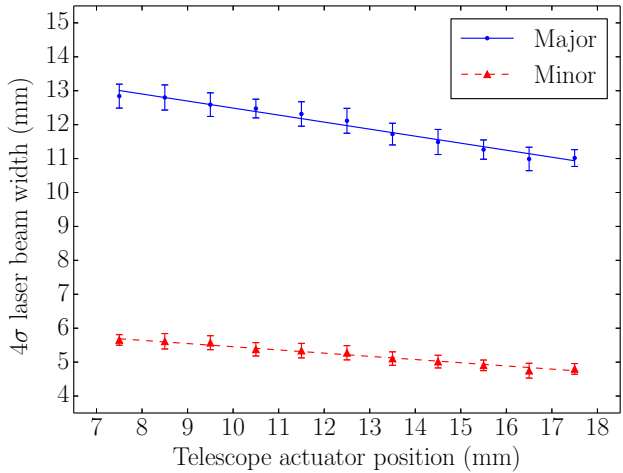}
\caption{\label{fig:ltelescopecalibration}$4\sigma$ widths of the major and minor 
axes of the input laser beam profile as measured at the end of the laser 
diagnostic line as a function of telescope actuator position.}
\end{figure}

The input beam size was adjusted using a demagnifying Gallilean telescope 
consisting of a plano-concave lens followed by a plano-convex lens that 
avoids ionisation when the laser beam is focussed in air. The telescope was placed 
close to the laser system aperture so that the telescope could be used to 
manipulate the input beam size without strongly affecting the laser divergence
and therefore the $x$ location of the laser focus at the LWIP. Additionally, this 
arrangement allows the insertion mirror for the diagnostic line to be placed 
after any optics that affect the laser divergence, ensuring the laser beam 
at the end of the diagnostic line is exactly the same as at the entrance 
to the laserwire lens.
A linear translation actuator was used to precisely adjust the spacing of the
lenses, minutely adjusting the divergence and therefore the input laser 
beam size at the laserwire lens. The $4\sigma$ laser beam widths (compliant 
with the \mbox{ISO 11145-2}~\cite{iso2005} standard) 
are shown in Figure~\ref{fig:ltelescopecalibration}. As the laser was
found to be astigmatic and elliptical, the major and minor beam widths of the 
beam ellipse are shown. The telescope actuator was set at 12.5\,mm for the 
laserwire operation period. 

\begin{figure}
\includegraphics[width=8.6cm]{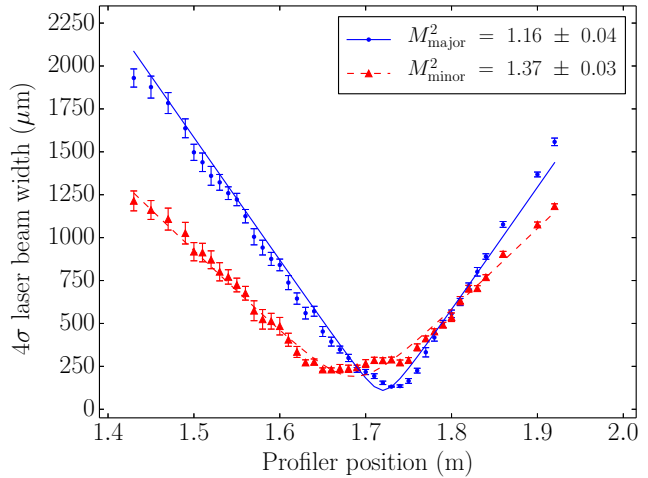}
\caption{\label{fig:lm2}Measured 4$\sigma$ widths of the laser beam through the focus 
created with a \mbox{$f~=~1.677$\,m} lens. The $M^{2}$ model is shown for each intrinsic axis of
the laser propagation, which were found to be rotated to the extrinsic lab axes by
-17.4$^{\circ}$.}
\end{figure}

The $M^{2}$ of the laser was measured by placing a 
\mbox{$f~=~1.677$\,m} (at $\lambda$~=~532\,nm) plano-convex lens at the end
of the laser diagnostic line to create a larger focussed spot size over a 
greater distance. Profiles of the laser beam were recorded at various positions 
throughout the focus. The $4\sigma$ widths along the intrinsic laser beam 
axes are shown in Figure~\ref{fig:lm2} along with a fit to the 
$M^{2}$ model (Equation~\ref{eq:lsigma} with $\Delta_{x}~=~0$).

\begin{figure}
\includegraphics[width=4.3cm]{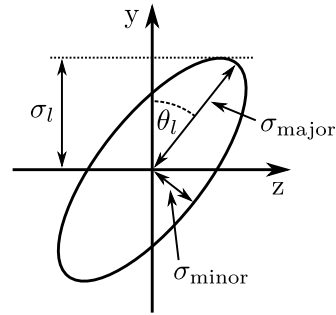}
\caption{\label{fig:gaussellipse} Maximum extent of an ellipse described by the major and 
minor axes $\sigma_{\rm{major}}$ and $\sigma_{\rm{minor}}$ respectively, here respresenting 
$\sigma_{y,z}$ of the bivariate Gaussian laser photon distribution.}
\end{figure}

\begin{figure}
\includegraphics[width=8.6cm]{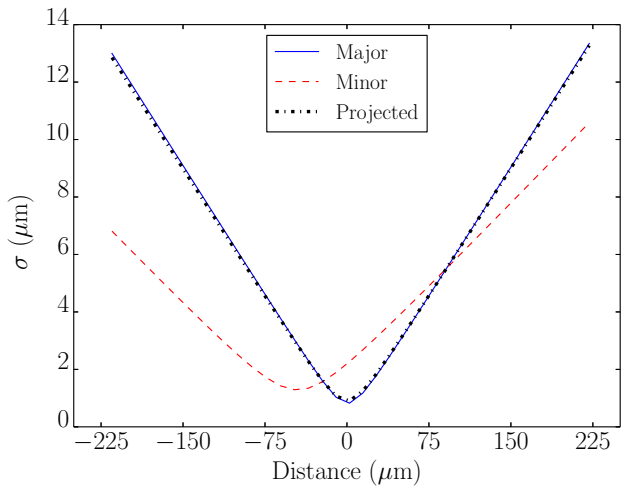}
\caption{\label{fig:lprojected}Calculated projected vertical sigma for the laser as well as the
two axes of propagation at the LWIP. The distance is zeroed about the minimum of the projected
vertical size where the laser is most intense and the Cherenkov signal greatest.}
\end{figure}

This shows that the laser is astigmatic with 
different focussed spot sizes at different locations with different divergences. The 
intrinsic axes of the laser were found to be rotated to the (extrinsic) lab axes by 
-17.4$^{\circ}$.
To deconvolve the laserwire scan, it is the distribution of photons in the vertical ($y$) 
axis that is required. To calculate this, the laser is assumed to be a 
bivarate Gaussian as described by $\sigma_{z,y}$. As the projection of a bivariate 
Gaussian distribution is also Gaussian, the relevant vertical projection 
is the maximum extent of the $\sigma_{l}$ ellipse depicted in Figure~\ref{fig:gaussellipse} 
and described by

\begin{equation}
\label{eq:lsigmaprojected}
\sigma_{l} = \sqrt{(\sigma_{lz}\,\sin\theta_{l})^{2} + (\sigma_{ly}\,\cos\theta_{l})^{2}}
\end{equation}

\noindent where $\theta_{l}$ is the angle of the laser axes with respect 
to the lab frame and the subscripts $z$ and $y$ denote the laser axis closest to that
dimension in the lab frame. Here, the major axis is closest to the $y$ dimension. 
The laser propagation parameters $\sigma_{o}$ and $x_{\sigma o}$ in each axis were scaled 
to the LWIP using the ratio of the focal lengths of the $M^{2}$ measurement lens and 
the laserwire 
lens. Each axis is described using Equation~\ref{eq:lsigma} using the scaled parameters
and the projected vertical size was calculated using Equation~\ref{eq:lsigmaprojected} 
as shown in Figure~\ref{fig:lprojected}. The laser propagation was measured each week 
after maintenance was carried out on the laser system and the relevant measurement used
in the analysis of the laserwire data.  In the case of the laserwire data presented here,
the minimum vertically projected laser spot size was \mbox{$\sigma_{l}~=~1.006~\pm~0.032$\,$\mu$m}.

The laser pointing stability was measured at the end of the laser diagnostic 
line to estimate the pointing stability of the laser at the LWIP by recording 
600 laser beam profiles and the centroid of each calculated. The standard 
deviation of the centroids in the horizontal and vertical was measured to be 125.7\,$\mu$m
and 132.7\,$\mu$m respectively at the laserwire lens. This measured spatial variation can
be scaled by the beam size at the ratio of the input laser beam size to that at the LWIP to 
give a laser position variation of $\sim$\,40\,nm in both dimensions. This spatial variation 
therefore systematically increases the measured electron beam size by approximately 0.08\,\%, 
which was deemed to be a negligible contribution and therefore not subtracted from the
laserwire scans.

\subsection{Electron Beam Characterisation}

At the start of laserwire operations, the laserwire electron beam optics were set
and the trajectory of the electron beam adjusted to quadrupole centres. After this, 
the dispersion and coupling were measured and corrected using the Flight Simulator 
software by changing the damping ring frequency in 1\,kHz steps over a range of 5\,kHz.
This was repeated several times to accurately correct coupling and dispersion. 
The measured residual dispersion at the LWIP was \mbox{$D(x)~=~4.215~\pm~0.515$\,mm} 
and \mbox{$D(y)~=~0.095~\pm~0.023$\,mm}.
This is acceptable and should make a negligible contribution to the vertical electron 
beam size, given the energy spread of the electron beam at the ATF2.

The emittance of the extracted electron beam in the ATF2 can be measured either using
wire-scanners or the multi-OTR system (mOTR)~\cite{Alabau2012}. Measuring the emittance 
using the wire-scanners during laserwire operations is impractical due to time
constraints. During early 2013, the mOTR system was being upgraded and was not available
for use during laserwire operations.

\subsection{Alignment\label{sec:alignment}}

To achieve collisions between the laser and electron beams, they must be spatially 
and temporally overlapped. Both of these functions were achieved using the OTR 
screen as an alignment tool. 

\subsubsection{Laser Alignment}

Before operations, the laser beam must be precisely aligned
to the centre of the laserwire lens as well as perpendicularly to the vacuum window
and lens assembly to ensure the diffraction limited focussed spot size is achieved. 
The low power alignment laser was first used without the laserwire
lens. The two mirrors before the LWIP were adjusted such that the back reflection from 
the vacuum window overlapped with the incoming laser beam back to its source. A mounted
mirror was then placed in the kinematic laserwire lens mount and the angle of the 
mount adjusted until the reflected laser beam also overlapped the incoming laser beam. 
This ensured the lens and 
window were parallel to each other and that no optical aberrations were introduced, as these 
would increase the focussed spot size. The alignment was 
verified with the main laser beam at low power. After this procedure, the mirror was removed 
from the lens mount and the laserwire lens was replaced.

\subsubsection{Spatial Alignment}

\begin{figure}[h]
\includegraphics[width=8.6cm]{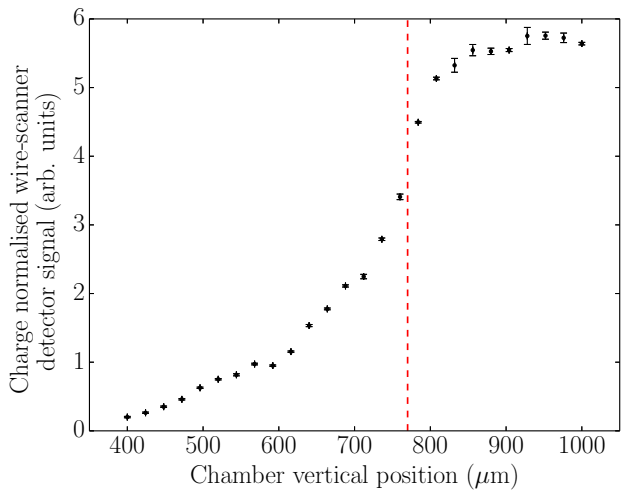}
\caption{\label{fig:verticalalignment}Measured brehmstraahlung radiation as a function 
of vertical chamber position with the OTR screen at the laser focus reference position. 
The red dashed line shows the chosen alignment position.}
\end{figure}

During access periods before operation, the laser was operated at low pulse energy and 
attenuated heavily so as not to cause damage to the OTR screen. The OTR screen was 
moved vertically to find the point where it intercepted the laser focus as observed 
in the post-LWIP optical system. The manual micrometers were adjusted to position the OTR 
screen along the $x$ axis so that the vertical distance required to occlude the laser beam
was minimised, ensuring that it was centred at the laser focus in 
the $x$ dimension. During experimental shifts, with the OTR screen set 
to the vertical reference position, the laserwire vacuum chamber was then scanned 
vertically until the electron beam was intercepted (the OTR screen arm  moves with 
the chamber). When the screen intercepts the electron beam brehmstraahlung radiation 
is produced that is detected by the wire-scanner detector behind the laserwire detector. 
The chamber was aligned to the point where half the maximum 
brehmstraahlung radiation was produced as shown in 
Figure~\ref{fig:verticalalignment}.

\begin{figure}[h]
\includegraphics[width=8.6cm]{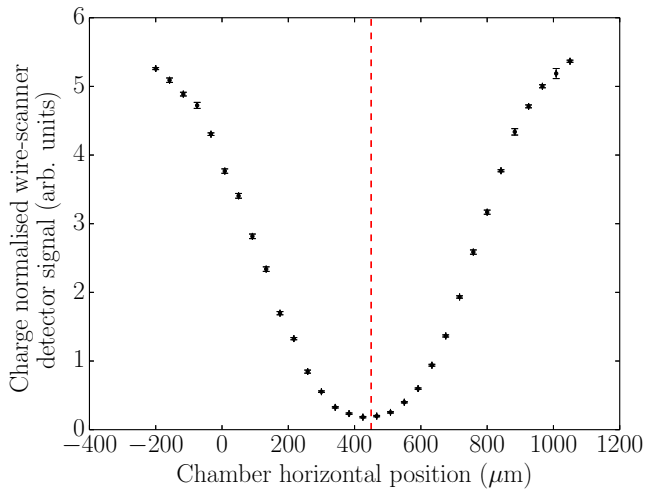}
\caption{\label{fig:horizontalalignment}Measured detected signal (brehmstraahlung) rate 
radiation as a function
of horizontal chamber position with the OTR screen at the laser focus reference position. 
The red dashed line shows the chosen alignment position.}
\end{figure}

During operations in 2011, the OTR screen was accidentally damaged by the high energy 
pulsed laser beam creating a semi-circular hole at the bottom of the screen 
approximately 500\,$\mu$m 
in diameter. This notch proved to be extremely useful as it allowed horizontal 
alignment of the laser beam for the first time. The OTR screen was placed 
approximately 200\,$\mu$m above the electron beam as found with vertical alignment and then 
the chamber was scanned in the horizontal $x$ axis. A minimum in brehmstraahlung 
radiation indicated the passage of the electron beam through the notch in the screen, 
which in turn indicates alignment to the laser focus. An example of this horizontal alignment 
scan is shown in Figure~\ref{fig:horizontalalignment}.

\subsubsection{Temporal Alignment}

To perform temporal alignment, the OTR screen was raised above the laser focus 
reference position to allow laser light to pass through the LWIP to the APD. The
vacuum chamber was then lowered until OTR was produced by the screen. 
The APD signal was observed on a 1\,GHz bandwidth, 5~giga-samples\,s$^{-1}$ oscilloscope. 
Due to bandwidth limitations of the APD, cables and the oscilloscope, both 
the laser and OTR pulses are represented by approximately 1\,ns pulses on the oscilloscope
used. However, it is still possible to perform the temporal alignment by 
attenuating the laser light
to match the OTR signal level and adjusting the laser system timing until the APD 
signal is doubled. This method allowed alignment within 40\,ps, which was sufficient
to attain detectable collisions.

\begin{figure}[h]
\includegraphics[width=8.6cm]{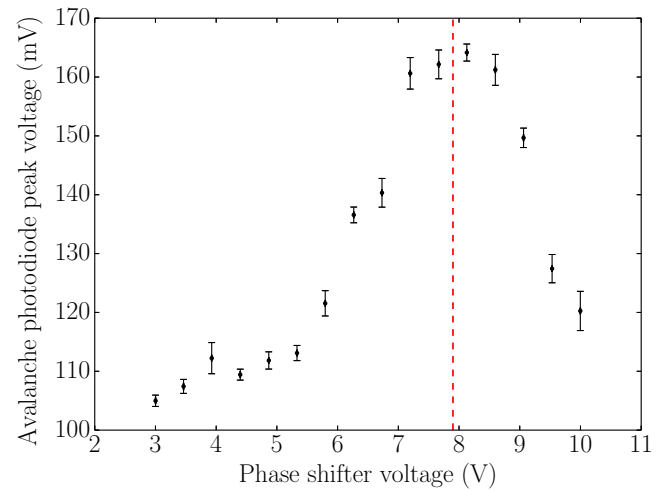}
\caption{\label{fig:timing}Peak APD signal as a function of laser phase controlled by a 
0~-~10\,V static voltage. The red dashed line shows the chosen phase setting.}
\end{figure}

The laser timing was intially adjusted in integer (357\,MHz, 2.8\,ns) clock cycles to align the 
laser and OTR signals as close as possible. The voltage-controlled phase shifter was then used
to adjust the phase between the electron beam and the laser pulses for precise alignment.
The maximum voltage of the oscilloscope trace was recorded as a function of 
phase shifter voltage as shown in Figure~\ref{fig:timing}, to ascertain the 
best timing overlap. Given the laser pulse and electron bunch lengths, the laser 
timing jitter of \textless~2\,ps as well as the electron bunch timing 
jitter of \textless~5\,ps, will increase the jitter of the detected laserwire signal
level by $\ll$~1\,\%, but will not affect the measured electron beam size.

A summary of the alignment accuracies in comparison to the nominal sizes in each dimension
is given in Table~\ref{tab:alignment}. This alignment procedure regularly led to detectable
collisions allowing optimisation of the alignment.

\begin{table}
\caption{\label{tab:alignment} Summary of the typical alignment accuracy $\Delta$ and 
the percentage of the nominal approximate size $\sigma$ in each dimension.}
\begin{tabular}{l r l r l c}
\hline \hline
Dimension\qquad\: & \multicolumn{2}{c}{$\Delta$} & \multicolumn{2}{c}{$\sigma$} & \% Error  \\ \hline
Vertical     & { }5 & $\mu$m  & { }1  &$\mu$m  & 500    \\
Phase        & 40   & ps      & 70  &ps        & { }57  \\
Horizontal   & 50   & $\mu$m  & 120 &$\mu$m    & { }41 \\ \hline\hline
\end{tabular}
\end{table}

\subsection{Collision Data}

\subsubsection{Initial Collisions}

Once detectable collisions were established, the alignment between the laser beam 
and the electron beam was optimised to produce the maximum number of Compton-scattered 
photons by scanning the laserwire vertically, then in phase, and then horizontally. 
This sequence of scans was repeated until no further improvement was observed.

During operations it was immediately clear that the vertical laserwire scans had 
a non-Gaussian shape with a narrow peak and broad wings observable up to 
30\,$\mu$m away from the peak of the scan. To accurately sample this shape in the minimum 
time, scans with variable step sizes were used, which were approximately distributed 
according to a cubic polynomial - here we call a \textit{nonlinear} scan. 
As a large number of steps are required in the 
centre of the scan, the centre of the scan must be within $\sim$\,1\,$\mu$m of the peak
of the measured signal for the best sampling. Therefore, an initial vertical scan with 
a low number of samples and linear step sizes was used for centring purposes. 
Furthermore, due to the non-Gaussian shape, Equation~\ref{eq:lwoverlapintegral} 
must be used. This requires knowledge of the horizontal electron beam size. 
To both optimise the alignment and measure the horizontal electron beam 
size, the laser focus was scanned horizontally across the electron beam 
over a 3\,mm range. After this, the chamber was positioned at the centre 
of the horizontal scan and a detailed nonlinear vertical scan 
was performed.

\subsubsection{Signal Linearity}

\begin{figure}
\includegraphics[width=8.6cm]{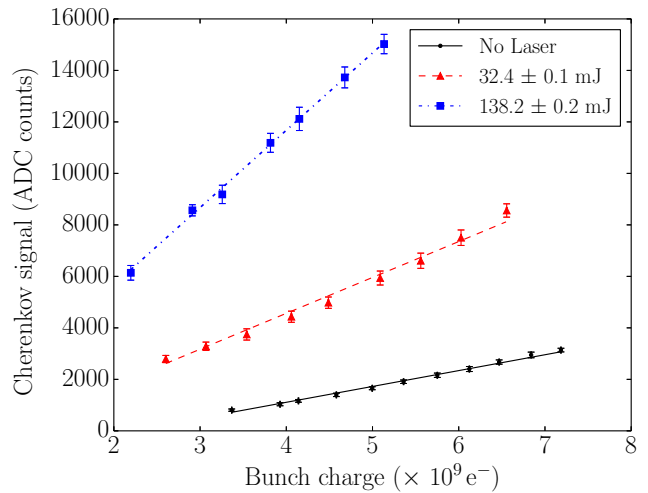}
\caption{\label{fig:chargecorrelations}Variation of the Cherenkov signal with electron
bunch charge for various laser levels.}
\end{figure}

With the Compton signal maximised, the electron bunch charge and laser pulse energy were
varied independently to ascertain the signal correlations and linearity. The bunch charge 
was varied by modulating the accelerator laser photocathode pulse energy. The delivered 
bunch charge to the extraction line varies very nonlinearly with photocathode laser
pulse energy, but the settings were chosen to give approximately linear steps. 
This charge ramp was repeated with no laser 
at the LWIP, as well as medium and maximum laser output power levels as shown in 
Figure~\ref{fig:chargecorrelations}.

\begin{figure}
\includegraphics[width=8.6cm]{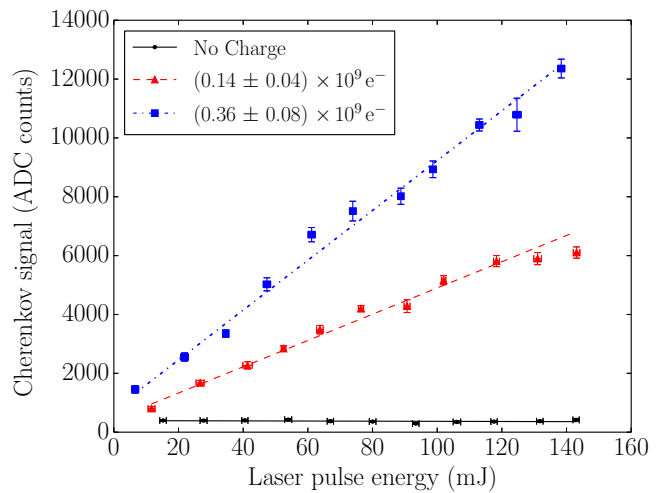}
\caption{\label{fig:lasercorrelations}Variation of the Cherenkov signal with laser pulse
energy for various electron bunch charges.}
\end{figure}

This shows a linear dependence with charge in all cases. The charge ramp was not continued 
to the highest bunch charge with the highest laser output to avoid detector 
saturation during the measurement. After this, a complementary set of scans were performed
by ramping the laser pulse energy while keeping the bunch charge fixed as shown
in Figure~\ref{fig:lasercorrelations}.

In this case, despite showing an approximately linear relationship with laser pulse 
energy as expected, a greater degree of variation was seen.
Here nonlinearities are most likely due to the variation of laser profile with laser 
pulse energy, which due to the laser technology employed is not expected to be 
consistent. The laser characterisation was carried out at the laser pulse
energy that would be used for operations.

These linearity scans encompass a much larger variation of both electron bunch charge and 
laser pulse energy than will be encountered during a scan and so from this data it can 
be seen that the detector will be linear over the small range of varation of bunch charge
and laser pulse energy during a laserwire scan.

\subsubsection{Combined Horizontal \& Vertical Analysis}

\begin{figure}
\includegraphics[width=8.6cm]{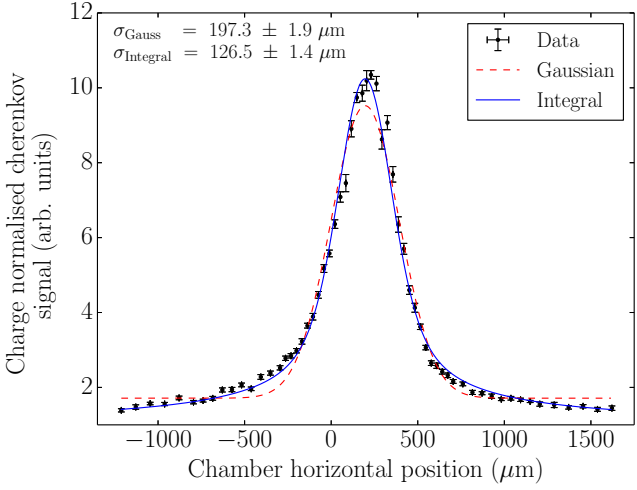}
\caption{\label{fig:hcomparison}Comparison of Gaussian and overlap integral  models for the horizontal laserwire scan.}
\end{figure}

The horizontal scans were initially fitted using a Gaussian model as this allows 
independent analysis of the horizontal and vertical scans. However, the horizontal scan is the 
convolution of the laser intensity in the $x$ axis with the Gaussian distribution of 
the electrons in the same dimension. In the case where the Rayleigh range is much 
less than the electron beam size 
\mbox{($x_{R}~\ll~\sigma_{ex}$)} the convolution is dominated by the electron beam shape and 
the Gaussian fit is acceptably accurate. Although the Gaussian model was found to 
provide an accurate fit in previous
operations~\cite{Boogert2010}, the horizontal scan data deviated from the 
Gaussian model curve noticeably.  
Therefore, the integral in Equation~\ref{eq:lwoverlapintegral} was used to 
fit a pair of horizontal and vertical scans
simultaneously to determine both $\sigma_{ex}$ and $\sigma_{ey}$. The horizontal scan
is shown in Figure~\ref{fig:hcomparison} with both the Gaussian and overlap integral
models for comparison.

Importantly, the extracted horizontal size is considerably different from that 
found using the Gaussian model, which if incorrectly used to deconvolve the vertical
laserwire scans yields an inaccurate vertical electron beam size. It had
originally been envisioned that a single horizontal scan could be used to 
deconvolve all the vertical laserwire scans for a given measurement period (such as an 8\,hour 
experimental shift). Even with adjustments made
to the vertical beam size that would affect the horizontal size, the deconvolution 
was expected to be relatively insensitive to the horizontal size.  However, even with 
changes in horizontal size of a few percent, this proved to be untenable and so 
horizontal and vertical scans were made each time for a complete measurement.

\subsubsection{Smallest Vertical Scan}

\begin{figure}
\includegraphics[width=8.6cm]{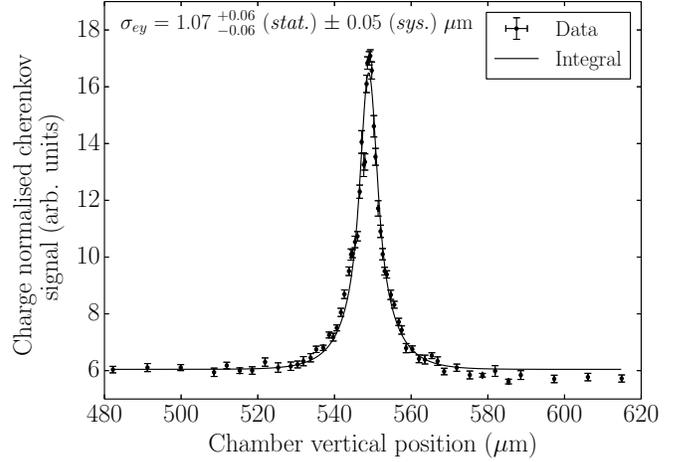}
\caption{\label{fig:nlvertical}Nonlinear step size laserwire scan with the smallest 
measured electron beam size.}
\end{figure}

\begin{figure}
\includegraphics[width=8.6cm]{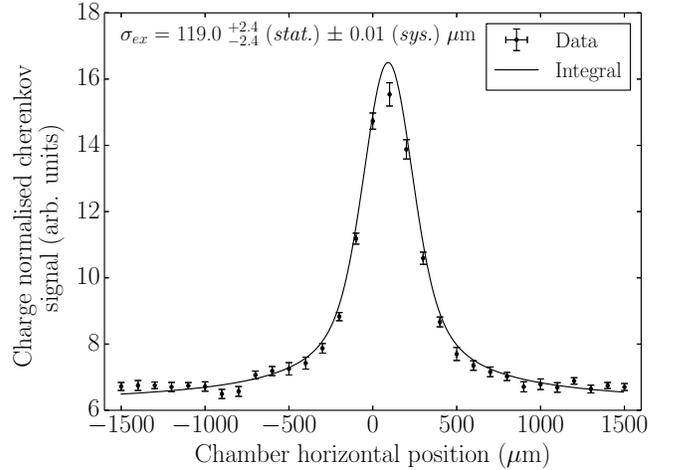}
\caption{\label{fig:horizontal}The corresponding horizontal laserwire scan for the smallest
vertical scan, which was required for the combined analysis.}
\end{figure}

The electron beam optics were manipulated to minimise the electron beam
size at the LWIP as measured by the laserwire. The laserwire scans shown in 
Figure~\ref{fig:nlvertical} and Figure~\ref{fig:horizontal} are the vertical and
horizontal laserwire scans respectively that were analysed together and 
constitute the smallest vertical electron beam profile measured. These were 
recorded with an electron bunch population of 
0.51~$\pm$~0.05\,$\times$\,10$^{10}$\,e$^{-}$.

\begin{figure}
\includegraphics[width=8.6cm]{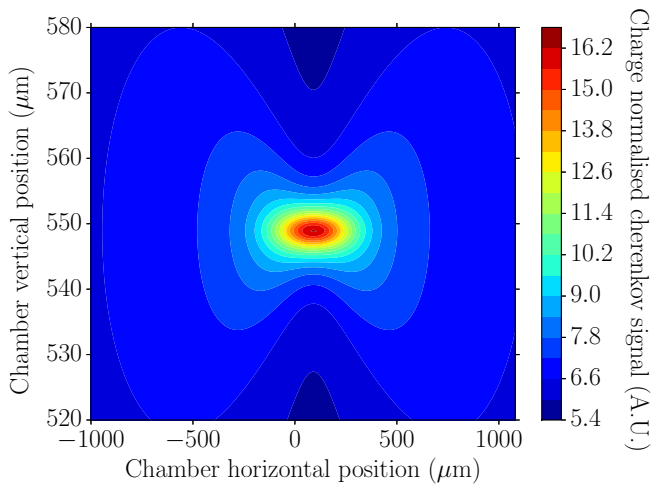}
\caption{\label{fig:doublefit}Calculated laserwire signal for a range of horizontal and
vertical chamber positions from the simultaneous fit of the horizontal and vertical laserwire 
scans.  The scans were performed immediately after each other ensuring consistent 
experimental conditions.}
\end{figure}

The measured vertical electron beam size was 
1.07~$^{+0.06}_{-0.06}$~(\textit{stat.})~$\pm$~0.05~(\textit{sys.})\,$\mu$m and
the horizontal beam size was 
119.0~$^{+2.4}_{-2.4}$~(\textit{stat.})~$\pm$~0.01~(\textit{sys.})\,$\mu$m. The 
analysis was performed using Minuit minimisation software using a weighted 
least squares method that allowed for asymmetrical
uncertainties using the Minos algorithm~\cite{James1975}. The systematic 
uncertainties were found by calculating the standard deviation of the fit parameters from
randomly sampling the laser parameters from the $M^{2}$ model analysis with their 
associated uncertainties. The calculated laserwire signal from the fit as a function 
of vertical and horizontal chamber positions is shown in Figure~\ref{fig:doublefit}. 
This shows that the vertical scan reaches a lower signal level than the horizontal 
scan at the edges of the scan, which can also be seen in 
Figures~\ref{fig:nlvertical}~and~\ref{fig:horizontal}.

\subsubsection{Quadrupole Scan}

\begin{figure}
\includegraphics[width=8.6cm]{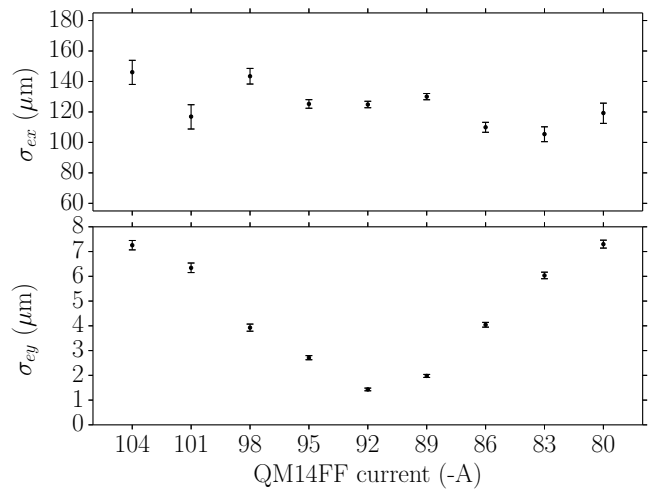}
\caption{\label{fig:quadscan}Horizontal (\textit{top}) and vertical (\textit{bottom}) electron 
beam sizes measured by the laserwire using combined analysis for various currents of 
QM14FF, the vertical focussing quadrupole immediately before the LWIP.}
\end{figure}

The laserwire was used to profile the electron beam throughout a quadrupole scan of the
vertically focussing quadrupole immediately before the LWIP, QM14FF. The magnet current was
varied from -80\,A to -104\,A in 3\,A steps. At each point, a short range, low sample number 
vertical scan was performed to vertically centre the laser beam. After this, a detailed 
horizontal scan
was performed followed by a nonlinear vertical scan. The horizontal and nonlinear vertical
scans were fitted together to yield $\sigma_{ex}$ and $\sigma_{ey}$ as shown in 
Figure~\ref{fig:quadscan}.

The vertical sizes show a clear hyperbolic focus as expected with a maximum measured size of 
7.30~$^{+0.16}_{-0.15}$~(\textit{stat.})~$\pm$~0.17~(\textit{sys.})\,$\mu$m and a smallest 
size of 1.43~$^{+0.05}_{-0.05}$~(\textit{stat.})~$\pm$~0.04~(\textit{sys.})\,$\mu$m. The 
horizontal sizes show a nearly linear progression as expected. The scan shows there is still 
clear variation in the measured scan size at $\sim$\,1.5\,$\mu$m with no flattening at 
the minimum of the scan, indicating the resolution limit of the laserwire has not been 
reached. The horizontal scan shows a greater degree of variation from the 
expected linear shape, which is due to variation in the electron beam energy over 
the time of the scans. The set of 3 laserwire scans required to make a 
measurement of the electron beam takes approximately 20\,mins and the whole quadrupole 
scan $\sim$\,3.5\,hours. A low electron bunch population of
0.2\,$\times$\,10$^{10}$\,e$^{-}$ was used in this case to provide the most stable 
condition over the duration of the measurement.

\begin{figure}
\includegraphics[width=8.6cm]{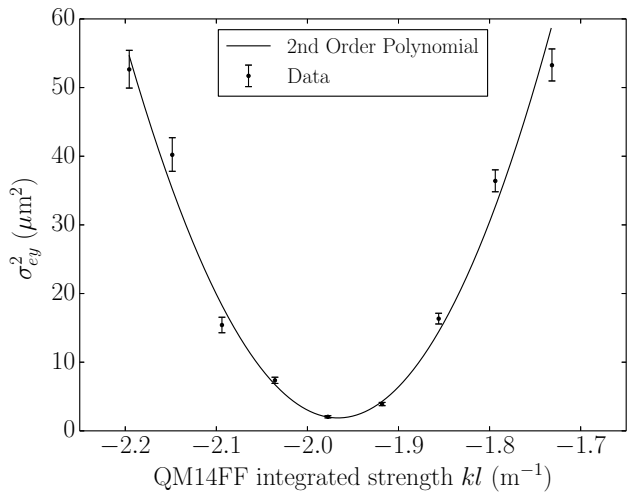}
\caption{\label{fig:quadfit}Measured electron beam size squared as a function of QM14FF 
strength. The least squares fit to a parabola for the emittance extraction is also shown.}
\end{figure}

Figure~\ref{fig:quadfit} shows the measured vertical sizes squared as a function of QM14FF
integrated quadrupole strength $kl$, which is modelled using the thick lens 
formalisim~\cite{Lee2004}

\begin{equation}
\label{eq:parabola}
\sigma_{ey}^{2} = a\,\left[\,m_{11}(kl) + b\,m_{12}(kl)\,\right]^{2} + c\,m_{12}^{2}(kl)
\end{equation}

\noindent where $a$, $b$ and $c$ are free parameters. $m_{11}$ and $m_{12}$ are given by

\begin{alignat}{1}
m_{11}(kl) &= S_{11}\cos\,(\sqrt{k}\,l) - S_{12}\sqrt{k}\sin\,(\sqrt{k}\,l)\label{eq:m11}
\\
m_{12}(kl) &= S_{11}\frac{1}{\sqrt{k}}\sin\,(\sqrt{k}\,l) - S_{12}\cos\,(\sqrt{k}\,l)\label{eq:m12}
\end{alignat}

\noindent where $S$ is the transfer matrix between the
quadrupole and the measurement plane. The geometric emittance is given by

\begin{equation}
\label{eq:emittance}
\epsilon = \sqrt{a\,c}
\end{equation}

There is only a drift segment of the beam line between the QM14FF quadrupole 
and the LWIP, so $S_{12}$ is the drift distance, \mbox{692.66~$\pm$~1.00\,mm} and 
$S_{11}$ is 1. 
From the fit to Equation~\ref{eq:parabola}, $a$ and $c$ were found to
be \mbox{2026.05~$\pm$~47.40\,$\times\,10^{-12}$\,m$^{2}$} and 
\mbox{3.365~$\pm$~0.235\,$\times$\,$10^{-12}$} 
respectively. Using these values, the measured projected geometric
emittance is \mbox{82.56~$\pm$~3.04\,pm\,rad}. This value is a relatively large 
emittance compared to the typical optimised vertical value of the ATF2, which in 
the damping ring is 10\,pm and 10\,-\,30\,pm in the extraction line.  The extraction 
line optics optimisation procedure~\cite{White2014} was not carried out in full 
for laserwire operation periods and higher emittance values are to be expected. 

\subsubsection{Detailed Vertical Slicing}

\begin{figure}
\includegraphics[width=8.6cm]{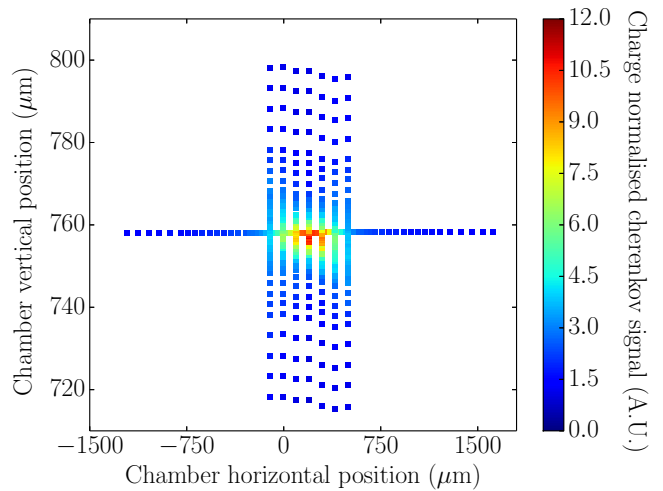}
\caption{\label{fig:multiscandata}Charge normalised chernekov signal sampled in multiple 
vertical laserwire scans at different horizontal positions. A long range horizontal scan is also included.}
\end{figure}

To fully map the laser-electron collisions, a detailed set of nonlinear
vertical laserwire scans was recorded at several horizontal locations as well as a long range
horizontal scan. All of the data was fitted simultaneously using the overlap integral model. 
The signal levels at each location are shown in 
Figure~\ref{fig:multiscandata}, and the parameters from the fit are given in 
Table~\ref{tab:multiscanfitparams}. In this case, the higher number of laser 
focus positions reduces the statistical uncertainty of the fit to the model significantly
and the systematic uncertainties now dominate. 

\begin{table}
\caption{\label{tab:multiscanfitparams}Parameters from fitting multiple laserwire scans. The first uncertainty 
is the asymmetric statistical uncertainty from the fit and the second is the associated systematic uncertainty.}
\begin{tabular}{l r@{}l l l}
\hline \hline
Parameter\quad\quad & \multicolumn{2}{c}{Fit value}& $\pm$ (\textit{stat.})& $\pm$ (\textit{sys.}) \\\hline \noalign{\vskip 1mm} 
Amplitude    &  9.&609   & $\pm~_{0.041}^{0.042}$  & $\pm$~0.049 \\
$x_{0}$       & 195.&728 & $\pm~_{0.525}^{0.525}$  & $\pm$~0.008 \\
$y_{0}$       & 756.&980 & $\pm~_{0.016}^{0.016}$ & $\pm$~0.002 \\
Background    & 1.&045  & $\pm~_{0.005}^{0.005}$   & $\pm$~0.021 \\
$\sigma_{ex}$ & 120.&588 & $\pm~_{0.650}^{0.651}$ & $\pm$~2.339 \\
$\sigma_{ey}$ & 1.&707   & $\pm~_{0.023}^{0.023}$ & $\pm$~0.051 \\\noalign{\vskip 1mm}
\hline\hline
\end{tabular}
\end{table}

Given the high aspect ratio of the electron beam, only a small 
amount of $x$-$y$ coupling would cause
the measured vertical projection of the electron beam to be significantly larger than the 
intrinsic vertical size of the electron beam.  Further more, there is the possibility that 
the laser beam could be at small angle with respect to the electron beam, which even if 
no coupling were present in the electron beam would result in a larger measured beam size.
Unlike the two scan analysis performed
already, the larger $x$-$y$ area covered by the data points allows the roll of the
electron beam to be analysed. This would conventionally be determined by applying a rotation 
of coordinates to the model and allowing the angle $\theta_{e}$ to be a free variable
in the minimisation. However, this would prevent an analytical solution to three dimensions
of the overlap integral, requiring a numerical solution for 
three dimensions instead of one. Although possible, this would
significantly complicate the data analysis and reduce the accuracy. Alternatively, 
if a small angle of rotation is assumed, the data
can be rotated instead of the model achieving approximately the same result. This can 
be safely assumed due to the high aspect ratio of the electron beam as a small rotation of
the data will not significantly affect the projection in the horizontal ($x$), whereas it 
will significantly affect the projection in the vertical ($y$).
If the electron beam is 
rotated with respect to the laserwire, the projection in the $y$ dimension, $\sigma_{ey'}$ 
will be measured and is described by Equation~\ref{eq:gaussprojection}. 

\begin{equation}
\label{eq:gaussprojection}
\sigma_{ey'} = \sqrt{(\sigma_{ey}\,\sin\theta_{e})^{2} + (\sigma_{ex}\,\cos\theta_{e})^{2}}
\end{equation}

\begin{figure}
\includegraphics[width=8.6cm]{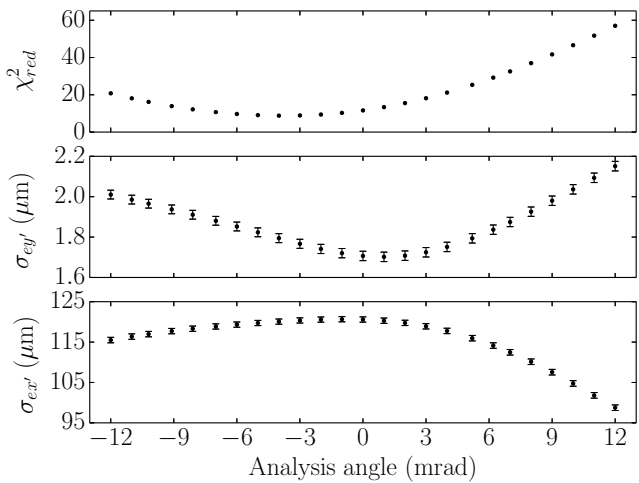}
\caption{\label{fig:multiscancomparison}Comparison of analysed vertical and horizontal sizes
from rotation of combined multiple laserwire scans as well as the reduced-$\chi^{2}$ for the
fit in each case.}
\end{figure}

As the angle of rotation $\theta_{e}$ increases, the measured vertical size increases 
quickly due to the high aspect ratio of the electron beam and an angle of 17\,m\,rad with
an aspect ratio of 100:1 for example would double the measured electron beam size. 
\mbox{$\sigma_{ey'}~=~\sigma_{ey}$} 
when the angle of analysis matches the angle of the electron beam with respect to the
laserwire. The reduced-$\chi^{2}$, $\sigma_{ey'}$ and $\sigma_{ex'}$ from fitting 
the data from Figure~\ref{fig:multiscandata} rotated by $\pm$~12\,mrad are shown in 
Figure~\ref{fig:multiscancomparison}.

\begin{figure}
\includegraphics[width=8.6cm]{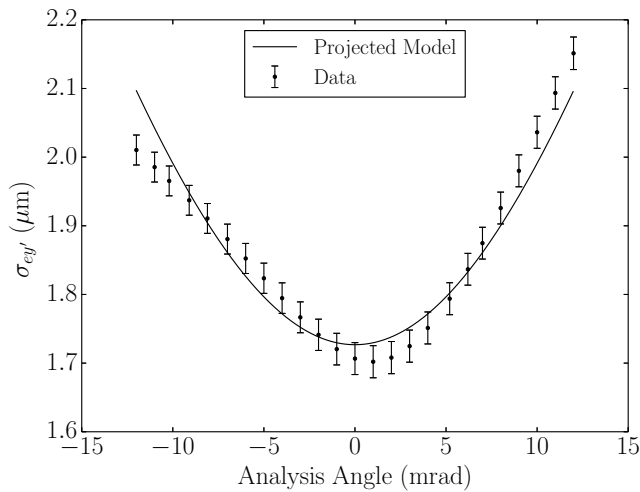}
\caption{\label{fig:multiscanverticalfit}Deconvolved vertical electron beam size
as a function of analysis angle for combined multiple laserwire scans. The fit to
a model of a projected bivariate Gaussian is also shown.}
\end{figure}

This shows a clear minimum with the smallest vertical size 
\mbox{$\sigma_{ey}~=~1.702~\pm~0.023$\,$\mu$m} at an angle of 1.0\,mrad. However,
$\sigma_{ex'}$ and $\sigma_{ey'}$ show a different maximum and mimum respectively 
and are also asymmetric about the minimum, which is not to be expected from the 
simple roll of the electron beam. This is shown more explicitly in 
Figure~\ref{fig:multiscanverticalfit}, where the model 
of Equation~\ref{eq:gaussprojection} is shown with $\sigma_{ey'}$. The fit of this model
indicates a minimum vertical size of
\mbox{$\sigma_{ey'}~=~1.727~\pm~0.007$\,$\mu$m} at an angle of 0.00~$\pm$~0.02\,mrad,
 but this model clearly does not accurately describe 
$\sigma_{ey'}\,(\theta_{e})$. 

Despite the minimum $\sigma_{ey'}$ and maximum $\sigma_{ex'}$ occuring at $\sim$\,0\,mrad, the 
minimum reduced-$\chi^{2}$ is found at a greater angle of $-4$\,mrad, indicating a better fit to
the data at this angle. These features are indicative of a further systematic effect not 
encapsulated by the model used. This is most likely due to the astigmatic propagation of the
laser beam. Although the $M^{2}$ used accurately describes the laser propagation in terms 
of the diameter of the laser beam, the transverse photon distribution is non-Gaussian 
and can be different on either side of the diffraction limited focus. This could be 
overcome with a laser with an improved $M^{2}$ and transverse profile.

\subsubsection{Measured \& Predicted Size}

During the laserwire operation period, the mOTR system was being upgraded and a comparative
emittance measurement was not possible. The measured electron 
beam sizes with the laserwire agree very well with predicted sizes from the MAD8 model
using the measured emittance by the laserwire and the measured dispersion.

The quadrupole scan made to measure the emittance using the laserwire was made at a 
different date than that of the smallest laserwire scan, however they agree well. The 
predicted size using Equation~\ref{eq:ebeamsize} is \mbox{1.129~$\pm$~0.021\,$\mu$m}
and the measured size was 
1.07~$^{+0.06}_{-0.06}$~(\textit{stat.})~$\pm$~0.05~(\textit{sys.})\,$\mu$m. 
The level of agreement shown is very good given that the optical functions at the 
extraction point in the damping ring were not measured during laserwire operations.
Furthermore, given the 3.12\,Hz repetition rate of the ATF2, further 
studies during one 8 hour operational shift are difficult.

\section{Outlook}

The laserwire system described in this paper has demonstrated a high resolution, 
effectively non-invasive method of measuring the transverse profile of a high charge
density, low emittance electron beam such as that at a future linear electron-positron 
collider. However, it
is clear that several improvements could be made to reduce the measurement uncertainty.

\subsection{Scanning Methods}

Whilst the scanning methodology employed that consisted of a single vertical and 
horizontal scan was
sufficient, it may not be optimal. From Figure~\ref{fig:doublefit}, it can be seen 
that a horizontal scan performed with a displacement of $\sim$\,15\,$\mu$m vertically 
would provide a scan with two maxima that may help to further constrain the fit to the model 
and 
reduce the uncertainty in the fitted parameters. Similarly, a coupled $x$-$y$ scan 
may provide further constraint. The nonlinear step size used in the scans was
highly effective and can be adjusted to the approximate size of the electron beam.

In the laserwire described in this paper, only one laser beam is used, 
whereas in an ideal laserwire diagnostic station there 
would be two laser beam lines orthogonal to each other; one for the vertical and 
one for the horizontal beam size measurement~\cite{Aumeyr2010}. In a similar 
case of an electron beam with
a high aspect ratio, the major axis of the elliptical beam could be measured
using a simple deconvolution requiring no knowledge of the minor axis size. This 
measurement of the major axis could then be used to deconvolve the laserwire scans of
the smaller minor axis that may exhibit Rayleigh range effects as described in this paper.
However, this would require knowledge of the precise horizontal offset of the laser focus with 
respect to the electron beam centre and so a further horizontal scan would be required.
As with the laserwire in this paper, it is possible to use a single laser beam to measure
both profiles, but only where the horizontal beam size is significantly greater than
the Rayleigh range of the laser beam. To this end, although one laser beam path may suffice,
two would guarantee laserwire measurements over a larger range of electron beam sizes and 
aspect ratios.

\subsection{Vacuum Chamber}

From the experiments conducted at the ATF2, it is possible to start the design process for a 
final ILC laserwire scanner. Our experimental vacuum chamber was engineered with 
flexibility and mechanical stability as the primary concerns, whilst a production ILC 
system will have to be optimised for cost and ease of operation. The authors would propose 
a smaller vacuum chamber, on a light weight mover system. Only a single station of ILC 
emittance measurement system would require an OTR system as the relative timing could 
be set from a single measurement for other nearby stations. The small laserwire vacuum chamber 
would be surrounded by a small optical system to move the laser beam onto the final focus 
lens and safely deal with the outgoing laser beam. Hopefully, in future the laser energy can be 
delivered to each station via a flexible fibre optic cable, without power loss and 
preserving the spatial and temporal quality of the beam. 

\subsection{Laser Source}

It is clear from the laser characterisation that while the laser source used provides
the necessary high peak power over a duration similar to the electron bunch 
length, the spatial quality could be improved. The laser source was astigmatic and 
had a non-Gaussian profile. In particular, the non-Gaussian nature makes the $M^{2}$
model an approximation of the size of the 
laser beam and not the true intensity distribution. A laser with a better spatial
profile would adhere much more closely to the $M^{2}$ model and reduce the systematic
uncertainty of the measured electron beam sizes~\cite{Saraceno2014}. 

Fibre lasers 
are a promising technology that could provide high spatial quality pulses at a 
much higher repetition rate with high overall efficiency. Recent work has focussed on 
achieving the necessary high peak powers from optical fibre lasers for such 
applications as a laserwire~\cite{Nevay2014}. Fibre lasers are not able to provide
the gigawatt peak powers used for this laserwire system, but in a well characterised
environment with a higher bunch charge electron beam such as a future linear collider,
the peak power requirements would be considerably lower~\cite{Agapov2007}.

\subsection{Intra-train Scanning}

For fast intra-train scanning the laser beam must be deflected in angle, unlike the 
mechanical motion of the lens and chamber used at the ATF2 laserwire system. 
The shift in position of the focussed laser spot is calculated using the deflection
angle $\theta$ and the appropriate transfer function of the optical path $f(\theta)$.
There are two main ways to angularly scan the laser beam; firstly, a piezo stack driven 
mirror~\cite{Aumeyr2010} or an electro-optic scanner~\cite{Bosco2009}. Alternatively, the 
possibility of dithering the position of the electron beam could be considered as the 
ranges vertically are in the 10s of micrometre ranges, although the beam rigidity could 
make this impractical and would rely on precise pulse to pulse beam position information.

\subsection{Laser Normalisation}

For a more complete system, the beam charge and laser pulse energy need to be measured 
with uncertainties significantly less than the Compton  or background signal variation. 
This is possible with high resolution direct current transformers for the particle beam, 
but the laser energy measurement systems would need more careful and systematic verification 
of performance before use with a laserwire system.

\subsection{BPMs for Laserwires}

The number and location of BPMs in the vicinity of laserwire stations will be important
in a multiple laserwire emittance measurement system. BPMs close to the laserwire interaction
points are useful for quick spatial and temporal alignment after initial collisions have 
been found. To enable spatial position jitter subtraction, high resolution
CBPMs will be required at each laserwire station. At the ATF2, the spatial position jitter
corresponded to $\sim$\,0.2\,$\sigma$, which therefore puts a stringent requirement of 
a resolution $\ll$~0.2\,$\mu$m to make a statistical difference. Such resolutions have
been demonstrated, but at the expense of dynamic range. Further study is required
to understand how well a full system of CBPMs can be used together with a laserwire.

\subsection{Signal \& Background Simulation}

For application of the laserwire to a high energy future linear collider, there will 
be many differences in the required setup for laserwire operations. In the 
case of a much higher energy electron beam, the Compton-scattered photons will 
receive a much larger fraction of the incoming electron energy, and it therefore may be
easier to measure the loss of the degraded energy electrons from the accelerator lattice than 
the Compton-scattered photons, or a combination of both. Simulations of the exact 
accelerator lattice will help determine this and also optimise the detector design. 

Maximising the signal to noise ratio of the Compton-scattering rate measurement is key
to improving the precision of the laserwire measurement and therefore minimising the 
error in the emittance measurement. Although the electron beam optics developed for laserwire
operations reduced the background significantly, the background signal level was still
significant and limits the precision of the measurement.
The effect of pointing jitter on the Compton-scattered 
photons as well as background conditions should be simulated and efforts to this end 
have been started using BDSIM~\cite{Deacon2008}.  

\section{Conclusions}

A laserwire capable of measuring high aspect ratio electron beams using 
a visible wavelength laser source has been demonstrated. A minimum vertical electron 
beam size of 1.07~$^{+0.06}_{-0.06}$~(\textit{stat.})~$\pm$~0.05~(\textit{sys.})\,$\mu$m 
was measured with a corresponding horizontal beam size was 
119.0~$^{+2.4}_{-2.4}$~(\textit{stat.})~$\pm$~0.01~(\textit{sys.})\,$\mu$m. A single
quadrupole scan was used to measure the vertically-projected geometric emittance of 
\mbox{82.56~$\pm$~3.04\,pm\,rad}.

Simultaneous fitting of the data from the horizontal and vertical laserwire
scans using the overlap integral model was 
demonstrated in the presence of Rayleigh range effects and was shown to provide 
an accurate measurement of both the horizontal and the vertical electron beam sizes.
This has demonstrated that Rayleigh range effects do not preclude the use of a 
laserwire to measure a high aspect ratio beam. Furthermore, an alignment method 
capable of finding collisions between the laser and electron beam in under 20~mins 
was developed. 

The laserwire was successfully operated with a low electron bunch population of
0.2\,$\times$\,10$^{10}$\,e$^{-}$ and will easily scale to 2\,$\times$\,10$^{10}$\,e$^{-}$.
We have identified a series of improvements and studies, which could improve this 
diagnostic, reduce systematic uncertainties and improve ease of engineering for a 
future laserwire.
Overall, a diagnostic for a future linear collider such as the ILC has 
been demonstrated, capable of measuring an electron beam size of 1\,$\mu$m.

\begin{acknowledgments}

We would like to thank the ATF2 collaborators and staff for their help in achieving
the necessary stable operating conditions to demonstrate the laserwire performance.
Additional thanks to the CBPM group, the OTR group and the SLAC team for 
many helpful discussions and help in characterising the machine. The research 
leading to these results has received funding from the Science and Technology 
Facilities Council via the John Adams Institute, Royal Holloway University of London, 
and the University of Oxford. We would like to acknowledge CERN for financial 
support of this research within CLIC-UK collaboration: Contract 
No KE1870/DG/CLIC as well as under the FP7 Research 
Infrastructures project Eu-CARD, grant agreement no. 227579.

\end{acknowledgments}

\bibliography{nevay2014lw}

\begin{thebibliography}{26}%
\makeatletter
\providecommand \@ifxundefined [1]{%
 \@ifx{#1\undefined}
}%
\providecommand \@ifnum [1]{%
 \ifnum #1\expandafter \@firstoftwo
 \else \expandafter \@secondoftwo
 \fi
}%
\providecommand \@ifx [1]{%
 \ifx #1\expandafter \@firstoftwo
 \else \expandafter \@secondoftwo
 \fi
}%
\providecommand \natexlab [1]{#1}%
\providecommand \enquote  [1]{``#1''}%
\providecommand \bibnamefont  [1]{#1}%
\providecommand \bibfnamefont [1]{#1}%
\providecommand \citenamefont [1]{#1}%
\providecommand \href@noop [0]{\@secondoftwo}%
\providecommand \href [0]{\begingroup \@sanitize@url \@href}%
\providecommand \@href[1]{\@@startlink{#1}\@@href}%
\providecommand \@@href[1]{\endgroup#1\@@endlink}%
\providecommand \@sanitize@url [0]{\catcode `\\12\catcode `\$12\catcode
  `\&12\catcode `\#12\catcode `\^12\catcode `\_12\catcode `\%12\relax}%
\providecommand \@@startlink[1]{}%
\providecommand \@@endlink[0]{}%
\providecommand \url  [0]{\begingroup\@sanitize@url \@url }%
\providecommand \@url [1]{\endgroup\@href {#1}{\urlprefix }}%
\providecommand \urlprefix  [0]{URL }%
\providecommand \Eprint [0]{\href }%
\providecommand \doibase [0]{http://dx.doi.org/}%
\providecommand \selectlanguage [0]{\@gobble}%
\providecommand \bibinfo  [0]{\@secondoftwo}%
\providecommand \bibfield  [0]{\@secondoftwo}%
\providecommand \translation [1]{[#1]}%
\providecommand \BibitemOpen [0]{}%
\providecommand \bibitemStop [0]{}%
\providecommand \bibitemNoStop [0]{.\EOS\space}%
\providecommand \EOS [0]{\spacefactor3000\relax}%
\providecommand \BibitemShut  [1]{\csname bibitem#1\endcsname}%
\let\auto@bib@innerbib\@empty
\bibitem [{\citenamefont {Aicheler}\ \emph {et~al.}(2012)\citenamefont
  {Aicheler}, \citenamefont {Burrows}, \citenamefont {Draper}, \citenamefont
  {Garvey}, \citenamefont {Lebrun}, \citenamefont {Peach},\ and\ \citenamefont
  {Phinney}}]{Clic2012}%
  \BibitemOpen
  \bibfield  {author} {\bibinfo {author} {\bibfnamefont {M.}~\bibnamefont
  {Aicheler}}, \bibinfo {author} {\bibfnamefont {P.}~\bibnamefont {Burrows}},
  \bibinfo {author} {\bibfnamefont {M.}~\bibnamefont {Draper}}, \bibinfo
  {author} {\bibfnamefont {T.}~\bibnamefont {Garvey}}, \bibinfo {author}
  {\bibfnamefont {P.}~\bibnamefont {Lebrun}}, \bibinfo {author} {\bibfnamefont
  {K.}~\bibnamefont {Peach}}, \ and\ \bibinfo {author} {\bibfnamefont
  {N.}~\bibnamefont {Phinney}},\ }\href@noop {} {\emph {\bibinfo {title} {A
  Multi-TeV Linear Collider Based on CLIC technology: CLIC Conceptual Design
  Report}}},\ \bibinfo {type} {Tech. Rep.}\ (\bibinfo {year}
  {2012})\BibitemShut {NoStop}%
\bibitem [{\citenamefont {Phinney}\ \emph {et~al.}(2007)\citenamefont
  {Phinney}, \citenamefont {Toge},\ and\ \citenamefont {Walker}}]{Ilc2007}%
  \BibitemOpen
  \bibfield  {author} {\bibinfo {author} {\bibfnamefont {N.}~\bibnamefont
  {Phinney}}, \bibinfo {author} {\bibfnamefont {N.}~\bibnamefont {Toge}}, \
  and\ \bibinfo {author} {\bibfnamefont {N.}~\bibnamefont {Walker}},\
  }\href@noop {} {\emph {\bibinfo {title} {International Linear Collider
  Reference Design Report Volume 3}}},\ \bibinfo {type} {Tech. Rep.}\ (\bibinfo
  {year} {2007})\BibitemShut {NoStop}%
\bibitem [{\citenamefont {Ross}\ \emph {et~al.}(2003)\citenamefont {Ross},
  \citenamefont {Anderson}, \citenamefont {Frisch}, \citenamefont {Jobe},
  \citenamefont {McCormick} \emph {et~al.}}]{Ross2002}%
  \BibitemOpen
  \bibfield  {author} {\bibinfo {author} {\bibfnamefont {M.}~\bibnamefont
  {Ross}}, \bibinfo {author} {\bibfnamefont {S.}~\bibnamefont {Anderson}},
  \bibinfo {author} {\bibfnamefont {J.}~\bibnamefont {Frisch}}, \bibinfo
  {author} {\bibfnamefont {K.}~\bibnamefont {Jobe}}, \bibinfo {author}
  {\bibfnamefont {D.}~\bibnamefont {McCormick}},  \emph {et~al.},\ }\href
  {\doibase 10.1063/1.1524406} {\bibfield  {journal} {\bibinfo  {journal} {AIP
  Conf. Proc.}\ }\textbf {\bibinfo {volume} {648}},\ \bibinfo {pages} {237}
  (\bibinfo {year} {2003})}\BibitemShut {NoStop}%
\bibitem [{\citenamefont {Hayano}(2000)}]{Hayano2000}%
  \BibitemOpen
  \bibfield  {author} {\bibinfo {author} {\bibfnamefont {H.}~\bibnamefont
  {Hayano}},\ }in\ \href@noop {} {\emph {\bibinfo {booktitle} {XX International
  Linac Conference}}}\ (\bibinfo {address} {Monterey, California},\ \bibinfo
  {year} {2000})\ pp.\ \bibinfo {pages} {146--148}\BibitemShut {NoStop}%
\bibitem [{\citenamefont {Tenenbaum}\ and\ \citenamefont
  {Shintake}(1999)}]{Tenenbaum1999}%
  \BibitemOpen
  \bibfield  {author} {\bibinfo {author} {\bibfnamefont {P.}~\bibnamefont
  {Tenenbaum}}\ and\ \bibinfo {author} {\bibfnamefont {T.}~\bibnamefont
  {Shintake}},\ }\href@noop {} {\bibfield  {journal} {\bibinfo  {journal}
  {Annual Review Nuclear Particle Physics}\ }\textbf {\bibinfo {volume} {49}},\
  \bibinfo {pages} {125} (\bibinfo {year} {1999})}\BibitemShut {NoStop}%
\bibitem [{\citenamefont {Agapov}\ \emph {et~al.}(2007)\citenamefont {Agapov},
  \citenamefont {Blair},\ and\ \citenamefont {Woodley}}]{Agapov2007}%
  \BibitemOpen
  \bibfield  {author} {\bibinfo {author} {\bibfnamefont {I.}~\bibnamefont
  {Agapov}}, \bibinfo {author} {\bibfnamefont {G.}~\bibnamefont {Blair}}, \
  and\ \bibinfo {author} {\bibfnamefont {M.}~\bibnamefont {Woodley}},\ }\href
  {\doibase 10.1103/PhysRevSTAB.10.112801} {\bibfield  {journal} {\bibinfo
  {journal} {Phys. Rev. ST Accel. Beams}\ }\textbf {\bibinfo {volume} {10}},\
  \bibinfo {pages} {112801} (\bibinfo {year} {2007})}\BibitemShut {NoStop}%
\bibitem [{\citenamefont {Ross}(2003)}]{Ross2003}%
  \BibitemOpen
  \bibfield  {author} {\bibinfo {author} {\bibfnamefont {M.}~\bibnamefont
  {Ross}},\ }in\ \href {\doibase 10.1109/PAC.2003.1288961} {\emph {\bibinfo
  {booktitle} {Particle Accelerator Conference}}}\ (\bibinfo {address}
  {Portland, USA},\ \bibinfo {year} {2003})\ pp.\ \bibinfo {pages}
  {503--507}\BibitemShut {NoStop}%
\bibitem [{\citenamefont {Alley}\ \emph {et~al.}(1996)\citenamefont {Alley}
  \emph {et~al.}}]{Alley1996}%
  \BibitemOpen
  \bibfield  {author} {\bibinfo {author} {\bibfnamefont {R.}~\bibnamefont
  {Alley}} \emph {et~al.},\ }\href {\doibase 10.1016/0168-9002(96)00556-6}
  {\bibfield  {journal} {\bibinfo  {journal} {Nucl. Instr. Methods A}\ }\textbf
  {\bibinfo {volume} {379}},\ \bibinfo {pages} {363 } (\bibinfo {year}
  {1996})}\BibitemShut {NoStop}%
\bibitem [{\citenamefont {Aumeyr}\ \emph {et~al.}(2010)\citenamefont {Aumeyr}
  \emph {et~al.}}]{Aumeyr2010}%
  \BibitemOpen
  \bibfield  {author} {\bibinfo {author} {\bibfnamefont {T.}~\bibnamefont
  {Aumeyr}} \emph {et~al.},\ }in\ \href@noop {} {\emph {\bibinfo {booktitle}
  {International Particle Accelerator Conference}}}\ (\bibinfo {address}
  {Kyoto, Japan},\ \bibinfo {year} {2010})\BibitemShut {NoStop}%
\bibitem [{\citenamefont {Boogert}\ \emph {et~al.}(2010)\citenamefont
  {Boogert}, \citenamefont {Blair}, \citenamefont {Boorman}, \citenamefont
  {Bosco}, \citenamefont {Deacon}, \citenamefont {Karataev}, \citenamefont
  {Aryshev}, \citenamefont {Fukuda}, \citenamefont {Terunuma}, \citenamefont
  {Urakawa}, \citenamefont {Corner}, \citenamefont {Delerue}, \citenamefont
  {Foster}, \citenamefont {Howell}, \citenamefont {Newman}, \citenamefont
  {Senanayake}, \citenamefont {Walczak},\ and\ \citenamefont
  {Ganaway}}]{Boogert2010}%
  \BibitemOpen
  \bibfield  {author} {\bibinfo {author} {\bibfnamefont {S.}~\bibnamefont
  {Boogert}}, \bibinfo {author} {\bibfnamefont {G.}~\bibnamefont {Blair}},
  \bibinfo {author} {\bibfnamefont {G.}~\bibnamefont {Boorman}}, \bibinfo
  {author} {\bibfnamefont {A.}~\bibnamefont {Bosco}}, \bibinfo {author}
  {\bibfnamefont {L.}~\bibnamefont {Deacon}}, \bibinfo {author} {\bibfnamefont
  {P.}~\bibnamefont {Karataev}}, \bibinfo {author} {\bibfnamefont
  {A.}~\bibnamefont {Aryshev}}, \bibinfo {author} {\bibfnamefont
  {M.}~\bibnamefont {Fukuda}}, \bibinfo {author} {\bibfnamefont
  {N.}~\bibnamefont {Terunuma}}, \bibinfo {author} {\bibfnamefont
  {J.}~\bibnamefont {Urakawa}}, \bibinfo {author} {\bibfnamefont
  {L.}~\bibnamefont {Corner}}, \bibinfo {author} {\bibfnamefont
  {N.}~\bibnamefont {Delerue}}, \bibinfo {author} {\bibfnamefont
  {B.}~\bibnamefont {Foster}}, \bibinfo {author} {\bibfnamefont
  {D.}~\bibnamefont {Howell}}, \bibinfo {author} {\bibfnamefont
  {M.}~\bibnamefont {Newman}}, \bibinfo {author} {\bibfnamefont
  {R.}~\bibnamefont {Senanayake}}, \bibinfo {author} {\bibfnamefont
  {R.}~\bibnamefont {Walczak}}, \ and\ \bibinfo {author} {\bibfnamefont
  {F.}~\bibnamefont {Ganaway}},\ }\href {\doibase
  10.1103/PhysRevSTAB.13.122801} {\bibfield  {journal} {\bibinfo  {journal}
  {Phys. Rev. ST Accel. Beams}\ }\textbf {\bibinfo {volume} {13}},\ \bibinfo
  {pages} {122801} (\bibinfo {year} {2010})}\BibitemShut {NoStop}%
\bibitem [{\citenamefont {Honda}\ \emph {et~al.}(2004)\citenamefont {Honda},
  \citenamefont {Kubo}, \citenamefont {Anderson}, \citenamefont {Araki},
  \citenamefont {Bane}, \citenamefont {Brachmann}, \citenamefont {Frisch} \emph
  {et~al.}}]{Honda2004}%
  \BibitemOpen
  \bibfield  {author} {\bibinfo {author} {\bibfnamefont {Y.}~\bibnamefont
  {Honda}}, \bibinfo {author} {\bibfnamefont {K.}~\bibnamefont {Kubo}},
  \bibinfo {author} {\bibfnamefont {S.}~\bibnamefont {Anderson}}, \bibinfo
  {author} {\bibfnamefont {S.}~\bibnamefont {Araki}}, \bibinfo {author}
  {\bibfnamefont {K.}~\bibnamefont {Bane}}, \bibinfo {author} {\bibfnamefont
  {A.}~\bibnamefont {Brachmann}}, \bibinfo {author} {\bibfnamefont
  {J.}~\bibnamefont {Frisch}},  \emph {et~al.},\ }\href {\doibase
  10.1103/PhysRevLett.92.054802} {\bibfield  {journal} {\bibinfo  {journal}
  {Phys. Rev. Lett.}\ }\textbf {\bibinfo {volume} {92}},\ \bibinfo {pages}
  {054802} (\bibinfo {year} {2004})}\BibitemShut {NoStop}%
\bibitem [{\citenamefont {{ATF2 Group}}(2005)}]{Report2006}%
  \BibitemOpen
  \bibfield  {author} {\bibinfo {author} {\bibnamefont {{ATF2 Group}}},\
  }\href@noop {} {\emph {\bibinfo {title} {ATF2 Proposal}}},\ \bibinfo {type}
  {Tech. Rep.}\ (\bibinfo  {institution} {KEK Report 2005-2},\ \bibinfo {year}
  {2005})\BibitemShut {NoStop}%
\bibitem [{\citenamefont {White}\ \emph {et~al.}(2014)\citenamefont {White}
  \emph {et~al.}}]{White2014}%
  \BibitemOpen
  \bibfield  {author} {\bibinfo {author} {\bibfnamefont {G.}~\bibnamefont
  {White}} \emph {et~al.},\ }\href {\doibase 10.1103/PhysRevLett.112.034802}
  {\bibfield  {journal} {\bibinfo  {journal} {Phys. Rev. Lett.}\ }\textbf
  {\bibinfo {volume} {112}},\ \bibinfo {pages} {034802} (\bibinfo {year}
  {2014})}\BibitemShut {NoStop}%
\bibitem [{\citenamefont {White}\ \emph {et~al.}(2008)\citenamefont {White},
  \citenamefont {Molloy}, \citenamefont {Seryi}, \citenamefont {Schulte},
  \citenamefont {Tomas}, \citenamefont {Kuroda}, \citenamefont {Bambade},\ and\
  \citenamefont {Renier}}]{White2008}%
  \BibitemOpen
  \bibfield  {author} {\bibinfo {author} {\bibfnamefont {G.}~\bibnamefont
  {White}}, \bibinfo {author} {\bibfnamefont {S.}~\bibnamefont {Molloy}},
  \bibinfo {author} {\bibfnamefont {A.}~\bibnamefont {Seryi}}, \bibinfo
  {author} {\bibfnamefont {D.}~\bibnamefont {Schulte}}, \bibinfo {author}
  {\bibfnamefont {R.}~\bibnamefont {Tomas}}, \bibinfo {author} {\bibfnamefont
  {S.}~\bibnamefont {Kuroda}}, \bibinfo {author} {\bibfnamefont
  {P.}~\bibnamefont {Bambade}}, \ and\ \bibinfo {author} {\bibfnamefont
  {Y.}~\bibnamefont {Renier}},\ }\href@noop {} {\emph {\bibinfo {title} {A
  Flight Simulator for ATF2: A Mechanism for International Collaboration in the
  Writing and Deployment of Online Beam Dynamics Algorithms}}},\ \bibinfo
  {type} {Tech. Rep.}\ (\bibinfo  {institution} {SLAC},\ \bibinfo {year}
  {2008})\BibitemShut {NoStop}%
\bibitem [{\citenamefont {Kim}\ \emph {et~al.}(2012)\citenamefont {Kim} \emph
  {et~al.}}]{Kim2012}%
  \BibitemOpen
  \bibfield  {author} {\bibinfo {author} {\bibfnamefont {Y.}~\bibnamefont
  {Kim}} \emph {et~al.},\ }\href {\doibase 10.1103/PhysRevSTAB.15.042801}
  {\bibfield  {journal} {\bibinfo  {journal} {Phys. Rev. ST Accel. Beams}\
  }\textbf {\bibinfo {volume} {15}},\ \bibinfo {pages} {042801} (\bibinfo
  {year} {2012})}\BibitemShut {NoStop}%
\bibitem [{\citenamefont {Johnston}(1998)}]{Johnston1998}%
  \BibitemOpen
  \bibfield  {author} {\bibinfo {author} {\bibfnamefont {T.}~\bibnamefont
  {Johnston}},\ }\href {http://www.ncbi.nlm.nih.gov/pubmed/18285945} {\bibfield
   {journal} {\bibinfo  {journal} {Appl. Opt.}\ }\textbf {\bibinfo {volume}
  {37}},\ \bibinfo {pages} {4840} (\bibinfo {year} {1998})}\BibitemShut
  {NoStop}%
\bibitem [{\citenamefont {Karataev}\ \emph {et~al.}(2011)\citenamefont
  {Karataev}, \citenamefont {Aryshev}, \citenamefont {Boogert}, \citenamefont
  {Howell}, \citenamefont {Terunuma},\ and\ \citenamefont
  {Urakawa}}]{Karataev2011}%
  \BibitemOpen
  \bibfield  {author} {\bibinfo {author} {\bibfnamefont {P.}~\bibnamefont
  {Karataev}}, \bibinfo {author} {\bibfnamefont {A.}~\bibnamefont {Aryshev}},
  \bibinfo {author} {\bibfnamefont {S.}~\bibnamefont {Boogert}}, \bibinfo
  {author} {\bibfnamefont {D.}~\bibnamefont {Howell}}, \bibinfo {author}
  {\bibfnamefont {N.}~\bibnamefont {Terunuma}}, \ and\ \bibinfo {author}
  {\bibfnamefont {J.}~\bibnamefont {Urakawa}},\ }\href {\doibase
  10.1103/PhysRevLett.107.174801} {\bibfield  {journal} {\bibinfo  {journal}
  {Phys. Rev. Lett.}\ }\textbf {\bibinfo {volume} {107}},\ \bibinfo {pages}
  {174801} (\bibinfo {year} {2011})}\BibitemShut {NoStop}%
\bibitem [{Epi(2014)}]{Epics2014}%
  \BibitemOpen
  \href {http://www.aps.anl.gov/epics/} {\enquote {\bibinfo {title}
  {Experimental physics and industrial control system
  \url{http://www.aps.anl.gov/epics/}},}\ } (\bibinfo {year}
  {2014})\BibitemShut {NoStop}%
\bibitem [{\citenamefont {Alabau-Gonzalvo}\ \emph {et~al.}(2012)\citenamefont
  {Alabau-Gonzalvo}, \citenamefont {Blanch~Gutierrez}, \citenamefont
  {Faus-Golfe}, \citenamefont {Garcia-Garrigos}, \citenamefont {Resta-Lopez}
  \emph {et~al.}}]{Alabau2012}%
  \BibitemOpen
  \bibfield  {author} {\bibinfo {author} {\bibfnamefont {J.}~\bibnamefont
  {Alabau-Gonzalvo}}, \bibinfo {author} {\bibfnamefont {C.}~\bibnamefont
  {Blanch~Gutierrez}}, \bibinfo {author} {\bibfnamefont {A.}~\bibnamefont
  {Faus-Golfe}}, \bibinfo {author} {\bibfnamefont {J.}~\bibnamefont
  {Garcia-Garrigos}}, \bibinfo {author} {\bibfnamefont {J.}~\bibnamefont
  {Resta-Lopez}},  \emph {et~al.},\ }in\ \href@noop {} {\emph {\bibinfo
  {booktitle} {International Particle Accelerator Conference}}}\ (\bibinfo
  {address} {San Sebasti\'{a}n, Spain},\ \bibinfo {year} {2012})\ pp.\ \bibinfo
  {pages} {879--881}\BibitemShut {NoStop}%
\bibitem [{iso(2005)}]{iso2005}%
  \BibitemOpen
  \href@noop {} {\enquote {\bibinfo {title} {{ISO 11146-2:2005 Lasers and
  laser-related equipment - Test methods for laser beam widths, divergence
  angles and beam propagation ratios - Part 2: General astigmatic beams}},}\ }
  (\bibinfo {year} {2005})\BibitemShut {NoStop}%
\bibitem [{\citenamefont {James}\ and\ \citenamefont {Roos}(1975)}]{James1975}%
  \BibitemOpen
  \bibfield  {author} {\bibinfo {author} {\bibfnamefont {F.}~\bibnamefont
  {James}}\ and\ \bibinfo {author} {\bibfnamefont {M.}~\bibnamefont {Roos}},\
  }\href {\doibase http://dx.doi.org/10.1016/0010-4655(75)90039-9} {\bibfield
  {journal} {\bibinfo  {journal} {Comput. Phys. Commun.}\ }\textbf {\bibinfo
  {volume} {10}},\ \bibinfo {pages} {343 } (\bibinfo {year}
  {1975})}\BibitemShut {NoStop}%
\bibitem [{\citenamefont {Lee}(2004)}]{Lee2004}%
  \BibitemOpen
  \bibfield  {author} {\bibinfo {author} {\bibfnamefont {S.~Y.}\ \bibnamefont
  {Lee}},\ }\href@noop {} {\emph {\bibinfo {title} {Accelerator Physics}}},\
  \bibinfo {edition} {2nd}\ ed.\ (\bibinfo  {publisher} {World Scientific},\
  \bibinfo {year} {2004})\BibitemShut {NoStop}%
\bibitem [{\citenamefont {Saraceno}\ \emph {et~al.}(2014)\citenamefont
  {Saraceno} \emph {et~al.}}]{Saraceno2014}%
  \BibitemOpen
  \bibfield  {author} {\bibinfo {author} {\bibfnamefont {C.~J.}\ \bibnamefont
  {Saraceno}} \emph {et~al.},\ }\href
  {http://www.ncbi.nlm.nih.gov/pubmed/24365808} {\bibfield  {journal} {\bibinfo
   {journal} {Opt. Lett.}\ }\textbf {\bibinfo {volume} {39}},\ \bibinfo {pages}
  {9} (\bibinfo {year} {2014})}\BibitemShut {NoStop}%
\bibitem [{\citenamefont {Nevay}\ \emph {et~al.}(2014)\citenamefont {Nevay},
  \citenamefont {Walczak},\ and\ \citenamefont {Corner}}]{Nevay2014}%
  \BibitemOpen
  \bibfield  {author} {\bibinfo {author} {\bibfnamefont {L.~J.}\ \bibnamefont
  {Nevay}}, \bibinfo {author} {\bibfnamefont {R.}~\bibnamefont {Walczak}}, \
  and\ \bibinfo {author} {\bibfnamefont {L.}~\bibnamefont {Corner}},\
  }\href@noop {} {\bibfield  {journal} {\bibinfo  {journal} {Phys. Rev. ST
  Accel. Beams (to be published)}\ } (\bibinfo {year} {2014})}\BibitemShut
  {NoStop}%
\bibitem [{\citenamefont {Bosco}\ \emph {et~al.}(2009)\citenamefont {Bosco},
  \citenamefont {Boogert}, \citenamefont {Boorman},\ and\ \citenamefont
  {Blair}}]{Bosco2009}%
  \BibitemOpen
  \bibfield  {author} {\bibinfo {author} {\bibfnamefont {A.}~\bibnamefont
  {Bosco}}, \bibinfo {author} {\bibfnamefont {S.}~\bibnamefont {Boogert}},
  \bibinfo {author} {\bibfnamefont {G.}~\bibnamefont {Boorman}}, \ and\
  \bibinfo {author} {\bibfnamefont {G.}~\bibnamefont {Blair}},\ }\href@noop {}
  {\bibfield  {journal} {\bibinfo  {journal} {Appl. Phys. Lett.}\ }\textbf
  {\bibinfo {volume} {94}} (\bibinfo {year} {2009})}\BibitemShut {NoStop}%
\bibitem [{\citenamefont {Deacon}\ and\ \citenamefont
  {Blair}(2008)}]{Deacon2008}%
  \BibitemOpen
  \bibfield  {author} {\bibinfo {author} {\bibfnamefont {L.}~\bibnamefont
  {Deacon}}\ and\ \bibinfo {author} {\bibfnamefont {G.}~\bibnamefont {Blair}},\
  }\href@noop {} {\bibfield  {journal} {\bibinfo  {journal} {EUROTeV}\ }\textbf
  {\bibinfo {volume} {018}} (\bibinfo {year} {2008})}\BibitemShut {NoStop}%
\end{thebibliography}%

\end{document}